\documentclass[aps,prl,nobibnotes,nofootinbib,showpacs, superscriptaddress, reprint]{revtex4-1}

\usepackage{graphicx}
\usepackage{amsmath} 	
\usepackage{xcolor}
\usepackage[utf8]{inputenc} 
\usepackage[T1]{fontenc}

\newcommand{\Table}[1]{Table~\ref{#1}} 
\newcommand{\ket}[1]{|#1\rangle}
\newcommand{\bra}[1]{\langle#1|}
\newcommand{\Fig}[1]{FIG.~\ref{#1}}
 
\newcommand{\Eq}[1]{Eq.~(\ref{#1})}

\newcommand{\wo}{\Omega_{0}}
\newcommand{\ko}{k_{0}}
\newcommand{\xo}{x_{0}}
\newcommand{\nnm}{N_{m}}

\newcommand{\kL}{k_{1}}
\newcommand{\kR}{k_{2}}
\newcommand{\wL}{\omega_{1}}
\newcommand{\wR}{\omega_{2}}
\newcommand{\wkd}{\omega_{_\textrm{KD}}}

\newcommand{\me}{m_{e}}
\newcommand{\qe}{q_{e}}

\newcommand{\tkd}{\tau_{_\textrm{KD}}}
\newcommand{\Ikd}{I_{_\textrm{KD}}}
\newcommand{\Lkd}{\lambda_{_\textrm{KD}}}

\newcommand{\Ltrap}{\lambda_{_\textrm{TL}}}
\newcommand{\Ttrap}{\tau_{_\textrm{TL}}}

\newcommand{\klaser}{k_{_\textrm{532}}}
\newcommand{\Itrap}{I_{_\textrm{TL}}}
\newcommand{\IS}{I_{_\textrm{S}}}
\newcommand{\deltap}{\delta_{p}}
\newcommand{\Dx}{\Delta x_{\textrm{cat}}}
\newcommand{\Dp}{\Delta p_{\textrm{cat}}}
\newcommand{\nmax}{n_{\textrm{max}}}
\newcommand{\nbk}{n_{\textrm{bk}}}

\newcommand{\wyTL}{W_{y\textrm{-TL}}}
\newcommand{\wzTL}{W_{z\textrm{-TL}}}
\newcommand{\wyKD}{W_{y\textrm{-KD}}}
\newcommand{\wzKD}{W_{z\textrm{-KD}}}

\begin{document}

\title{Kapitza-Dirac blockade: A universal tool for the \\ deterministic preparation of non-Gaussian oscillator states}

\author{Wayne~Cheng-Wei~Huang}
\email{Email: waynehuang1984@gmail.com \\ Current address: University of G\"{o}ttingen, IV. Physical Institute, 370077 G\"{o}ttingen, Germany.} 
\affiliation{Department of Physics and Astronomy, Northwestern University, Evanston, Illinois 60208, USA}

\author{Herman~Batelaan}
\affiliation{Department of Physics and Astronomy, University of Nebraska-Lincoln, Lincoln, Nebraska 68588, USA}

\author{Markus~Arndt}
\affiliation{Faculty of Physics, University of Vienna, Boltzmanngasse 5, A-1090 Vienna, Austria}

\begin{abstract}
Harmonic oscillators count among the most fundamental quantum systems with important applications in molecular physics, nanoparticle trapping, and quantum information processing. 
Their equidistant energy level spacing is often a desired feature, but at the same time a challenge if the goal is to deterministically populate specific eigenstates.
Here, we show how interference in the transition amplitudes in a bichromatic laser field can suppress the sequential climbing of harmonic oscillator states (Kapitza-Dirac blockade) and achieve selective excitation of energy eigenstates, Schr\"{o}dinger cats and other non-Gaussian states.
This technique can transform the harmonic oscillator into a coherent two-level system or be used to build a large-momentum-transfer beam splitter for matter-waves. To illustrate the universality of the concept, we discuss feasible experiments that cover many orders of magnitude in mass, from single electrons over large molecules to dielectric nanoparticles.
\end{abstract}

\maketitle 
The harmonic oscillator is a paradigmatic text book example of fundamental quantum physics and it has remained at the heart of modern research. Quantum harmonic oscillators have been realized with single electrons \cite{Brown1986, Hanneke2008}, single ions \cite{Leibried2003}, ultra-cold quantum gases \cite{Cornell2001}, and dielectric nanoparticles \cite{Delic2020}. For all these systems, cooling to the oscillator ground state has been successfully demonstrated. Our present proposal is motivated by the challenge to prepare highly non-classical states, mesoscopic Schr\"{o}dinger cat states and large-momentum transfer beam splitters, independent of detailed oscillator properties. 

Throughout the last two decades, macroscopic quantum superposition was realized in widely different systems \cite{Arndt2014}. Neutrons were delocalized over 10\,cm \cite{Zawisky2002}. Atoms were put in superpositions on the half-meter scale \cite{Kovachy2015} or in momentum states separated by more than 1000\,$\hbar k$ \cite{Gebbe2019}, and molecules in excess of 25,000 Da were delocalized over hundred times their size \cite{Fein2019}. Lately, it has been proposed to prepare dielectric nanoparticles in distinct position states \cite{Arndt2014, Bateman2014} to test the nature of quantum collapse \cite{Bassi2013}, quantum decoherence \cite{Joos1985, Wootters1979, Zurek1991}, or even the quantum nature of gravity \cite{Bose2017, Marletto2017}. 

Here we propose to exploit Kapitza-Dirac blockade as a universal tool for preparing Schr\"{o}dinger cat or non-Gaussian states in general, with single electrons, molecules and nanoparticles, differing in mass by more than nine orders of magnitude.

\begin{figure}[b]
\centering
\scalebox{0.48}{\includegraphics{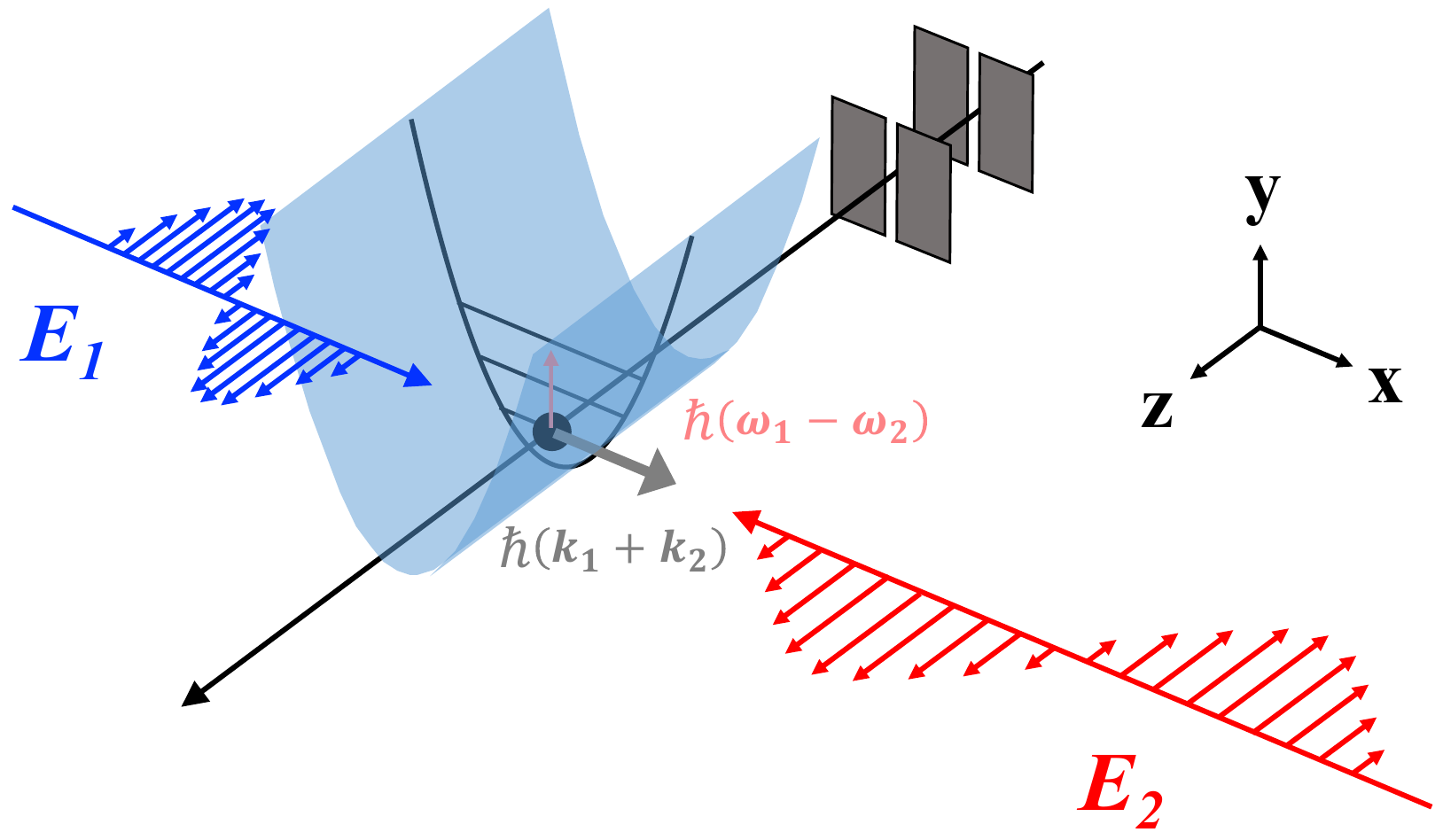}}
\caption{(color online) Proposed realization of the Kapitza-Dirac blockade and quantum state control in a 1D harmonic oscillator. A collimated particle populates the ground state of a 1D harmonic trap. A pair of counter-propagating bichromatic KD-laser fields $E_{1,2}$ interacts with the particle and changes its energy and momentum while it is in the trap.}
\label{fig:schematic}
\end{figure}

In the original proposal by Kapitza and Dirac \cite{Kapitza1933}, electrons were assumed to be coherently scattered by the ponderomotive force of a standing light wave. It took several advances in electron beam and laser technology to realize this idea 70 years later, both in the Raman-Nath and in the Bragg regime \cite{Freimund2001, Freimund2002}. The idea of optical phase gratings has also been extended to atoms \cite{Gould1986, Martin1988, Pfau1994} and molecules \cite{Nairz2001, Brand2020}, where the interaction is mediated by the optical dipole force between the laser electric field and the particle's polarizability. Recently, the Kapitza-Dirac effect has been used in electron microscopy to facilitate all-optical phase masks \cite{Talebi, Schwartz}. 

The \emph{inelastic} Kapitza-Dirac (KD) effect was first discussed in \cite{Huang2019}. It differs from its elastic counterpart not only by the use of laser fields with different frequencies, but also by the presence of a harmonic trap. The latter modifies the conditions for energy and momentum conservation. The inelastic KD-effect is similar to stimulated Raman scattering \cite{Kasevich1991, Kozuma1999, Hemmerich1994, Monroe1995} but operates without internal states, so that absorption, spontaneous emission and decoherence in objects with broad resonance lines can be avoided. It can thus be applied to particles that do not exhibit any internal states at all. 

In this letter, we introduce Kapitza-Dirac blockade as a new feature emerging from the quantum mechanical treatment of the inelastic KD-effect. By using judiciously chosen laser frequencies, the Kapitza-Dirac blockade can drive a controlled parametric resonance while blocking undesired transitions almost entirely. 

We start for simplicity with a particle traveling through a 1D harmonic trap (see \Fig{fig:schematic}). The trap can be realized via the ponderomotive potential on electrons or via the optical dipole potential on polarizable particles. The scheme preferably starts from the harmonic oscillator ground state, which can be populated with high probability by filling the trap with a tightly collimated particle beam.  

The two KD-laser pulses with frequencies $\omega_{1,2}$ ($\wL > \wR$) and a $1/e$ pulse duration $\tkd$ are assumed to propagate counter to each other along the $x$-axis. Their polarization should be chosen to avoid wave-mixing with the trapping lasers. 

\begin{figure}[t]
\centering
\scalebox{0.59}{\includegraphics{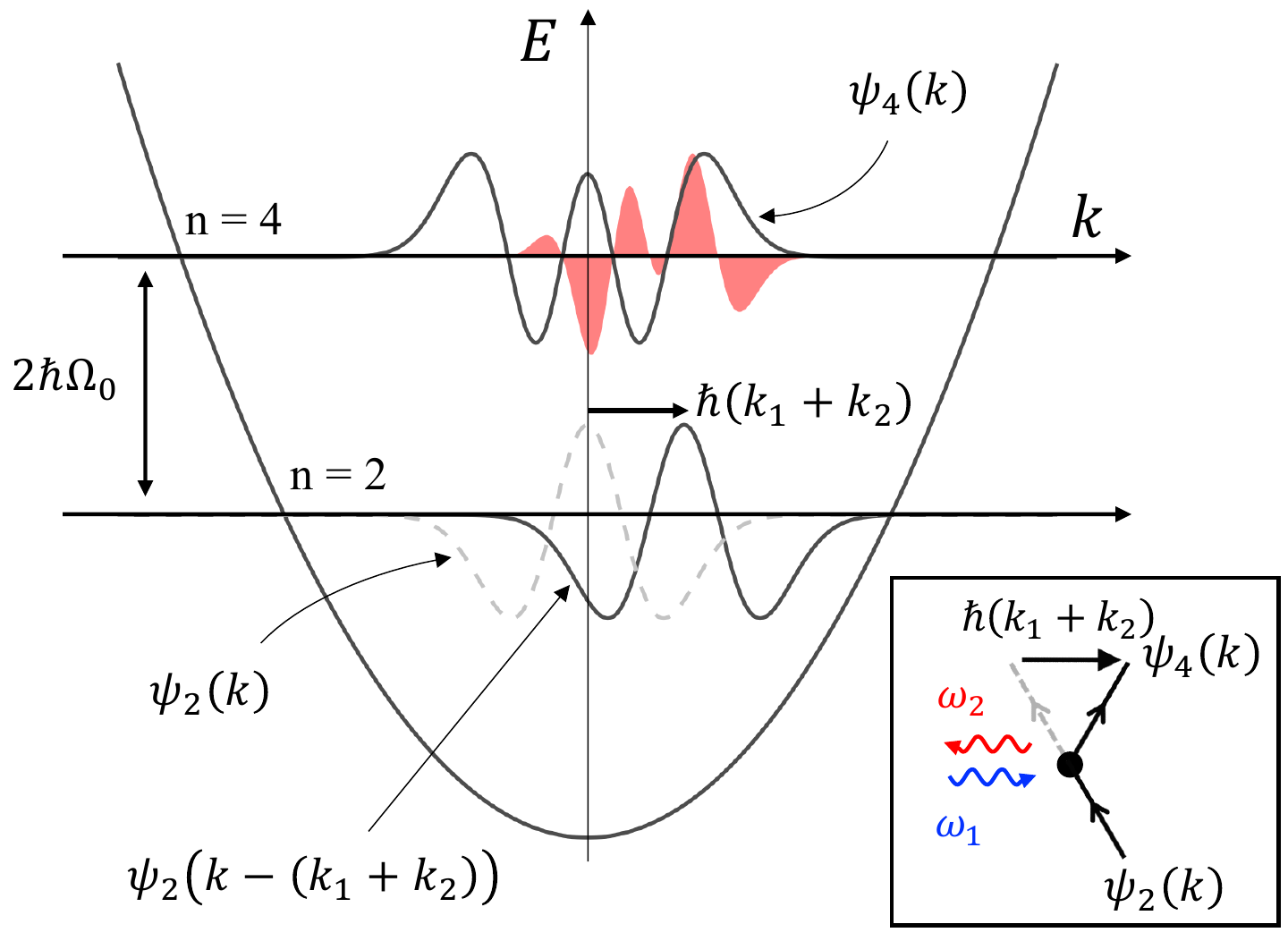}}
\caption{(color online) Kapitza-Dirac blockade in the harmonic oscillator for the transition between the energy eigenstates $\ket{n=2} \rightarrow \ket{n=4}$. Energy and momentum conservation require an energy change of $\hbar(\wL - \wR) = 2\hbar\Omega_0$ and a momentum recoil of $\hbar(\kL + \kR) = \left( 2 + \deltap \right) \hbar\ko$ (inset). When the overlap integral vanishes, i.e. for  $g_{2}(\eta)\propto \int_{-\infty}^\infty \psi^{\ast}_{4}(k) \psi_{2}(k- (\kL + \kR)) dk =0$, this transition is suppressed, even in the presence of resonant laser light. The red-shaded integrand  $\psi^{\ast}_{4}(k) \psi_{2}(k-(\kL + \kR))$   represents the weight of all vertical transition amplitudes starting from different k-values. Momentum needs to be tuned for destructive interference to null the overlap integral. The potential is shown with vertical compression for clearer illustration of the wavefunctions.}
\label{fig:interfere}
\end{figure}

The oscillator can be parametrically driven through an effective Hamiltonian \cite{Suppl}
\begin{equation}
	\hat{H}_{int} = \mathcal{C} E_{1}(t)E_{2}(t) \cos{\left( (\kL+\kR)x -(\wL-\wR)t \right)},
\end{equation}
where $\mathcal{C}$ is the system specific coupling coefficient, $k_{1,2} = \omega_{1,2}/c$, and $E_{1,2}(t) = E_{1,2}\exp{(-t^2/\tkd^2)}$ is the time envelope of the KD-laser pulse. Resonance occurs when 
\begin{equation}\label{conservation}
	\left\{
	\begin{array}{l}
		\displaystyle	\wL - \wR = \nnm \wo \\	
		\displaystyle	\kL + \kR = \left( \nnm + \deltap \right) \ko	,
	\end{array}	
	\right.	
\end{equation}
where $\wo$ is the harmonic trap frequency, $\nnm$ is a positive integer and $\deltap \ge -\nnm$. The dimensionless momentum detuning $\deltap$ characterizes the distance of $\hbar(\kL + \kR)$ to its maximal value $\nnm p_0=\nnm\hbar\ko$ before the overlap integral between the momentum wavefunctions decreases (see \Fig{fig:interfere}). Here $p_0= \sqrt{\hbar m \wo/2}$ is the standard deviation of the oscillator's ground-state momentum probability distribution. 

The resonant KD-laser frequencies can be found from \Eq{conservation} to be $\omega_{1,2} = \wkd \pm \nnm\wo/2$, where  the central KD-frequency is
\begin{equation}\label{KDfreq}
	\wkd \equiv (\nnm+\deltap)c\ko/2.
\end{equation} 
To demonstrate the preparation of non-Gaussian harmonic oscillator states, we choose $\nnm = 2$ and evaluate the dimensionless transition amplitude $g_{n} (\eta) \equiv \bra{n+2}\cos{\left( (\kL+\kR)x \right)}\ket{n}$ for the transition from an oscillator energy eigenstate $\ket{n}$ to $\ket{n+2}$ \cite{Suppl, Cahill1969, Wineland1998}, 
\begin{equation}\label{transition}
	g_{n} (\eta) = -\sqrt{\frac{n!}{(n+2)!}} \eta^{2} L^{(2)}_{n} (\eta^2) e^{-\eta^2/2}.
\end{equation}
Here $L^{(2)}_{n}(y)$ is the generalized Laguerre polynomial. The Lamb-Dicke parameter is defined as $\eta \equiv (\kL+\kR)\xo$, where $\xo = \sqrt{\hbar/2 m \wo}$ is the standard deviation of the oscillator's ground-state position probability distribution.

\begin{figure*}[t]
\centering
\scalebox{0.6}{\includegraphics{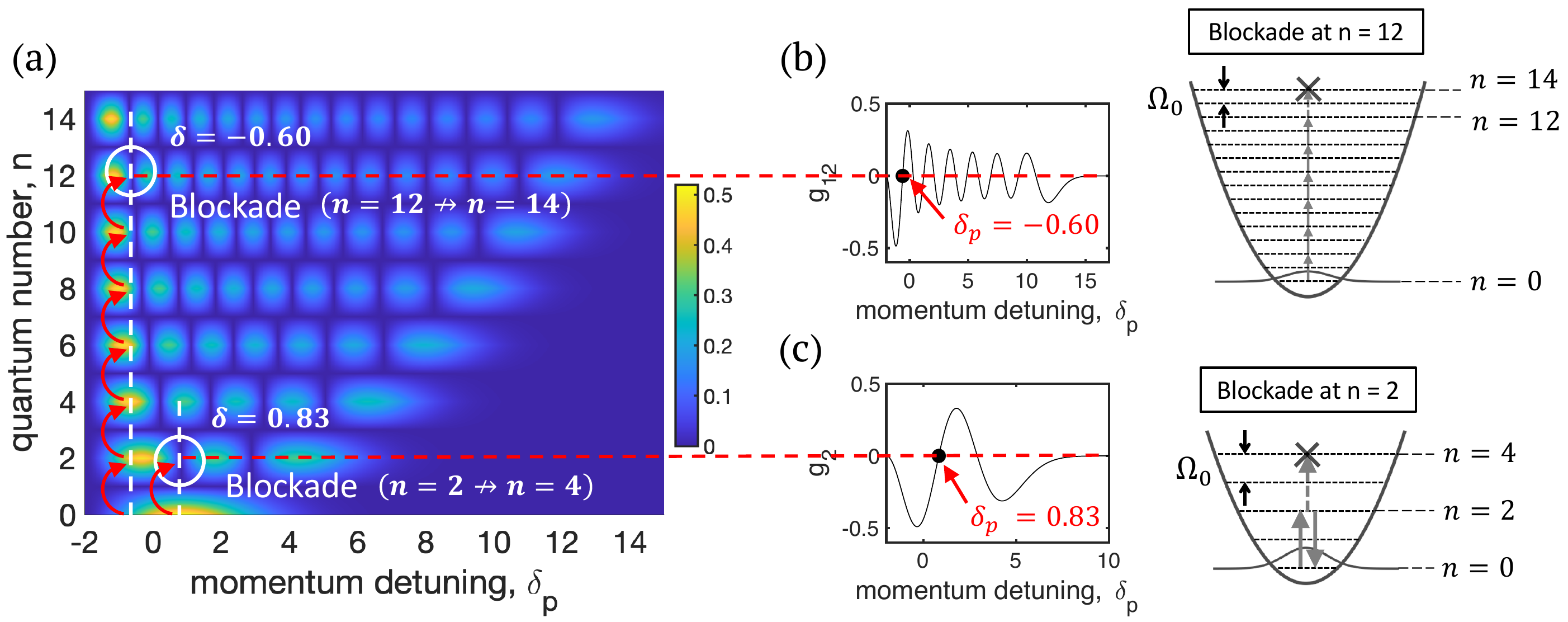}}
\caption{(color online) Transition map for the inelastic KD-effect. (a) The norm of the transition amplitude $|g_{n}|$ in \Eq{transition} is plotted as a function of the quantum number $n$ and the momentum detuning $\deltap$. The nodes along the $\deltap$-axis (white circles) can be used to stop sequential excitations (red curved arrows). The width along the $n$-axis does not illustrate the oscillator linewidths but is chosen for clearer illustration of the energy levels. (b) The zero-crossing of $g_{12}(\eta)$ at $\deltap = -0.60$ is used to stop transitions beyond $\ket{n=12}$. (c) The zero-crossing $g_{2}(\eta)$ at $\deltap = 0.83$ is used to prepare an effective two-level system between $\ket{n=0}$ and $\ket{n=2}$.}
\label{fig:transition_map}
\end{figure*}

The transition amplitude $g_{n} (\eta)$ has zero-crossings (see \Fig{fig:transition_map}(a)), which implies that the transition $\ket{n} \rightarrow \ket{n+2}$ can be suppressed at certain values of the momentum detuning $\deltap$ because of destructive interference between transition amplitudes starting from different k-values  (see \Fig{fig:interfere}). This \emph{Kapitza-Dirac blockade} is a powerful tool, as it can stop the sequential excitation in the energy ladder and allows us to prepare non-Gaussian harmonic oscillator states. The energy-momentum conservation \Eq{conservation} implies that the Lamb-Dicke parameter  $\eta = (\nnm + \deltap)/2$ is independent of any oscillator details. In consequence, the Kapitza-Dirac blockade is independent of the specific oscillator realization, and only the central KD-frequency $\wkd = c\ko \eta$ is oscillator dependent. 

\begin{figure}[t]
\centering
\scalebox{0.42}{\includegraphics{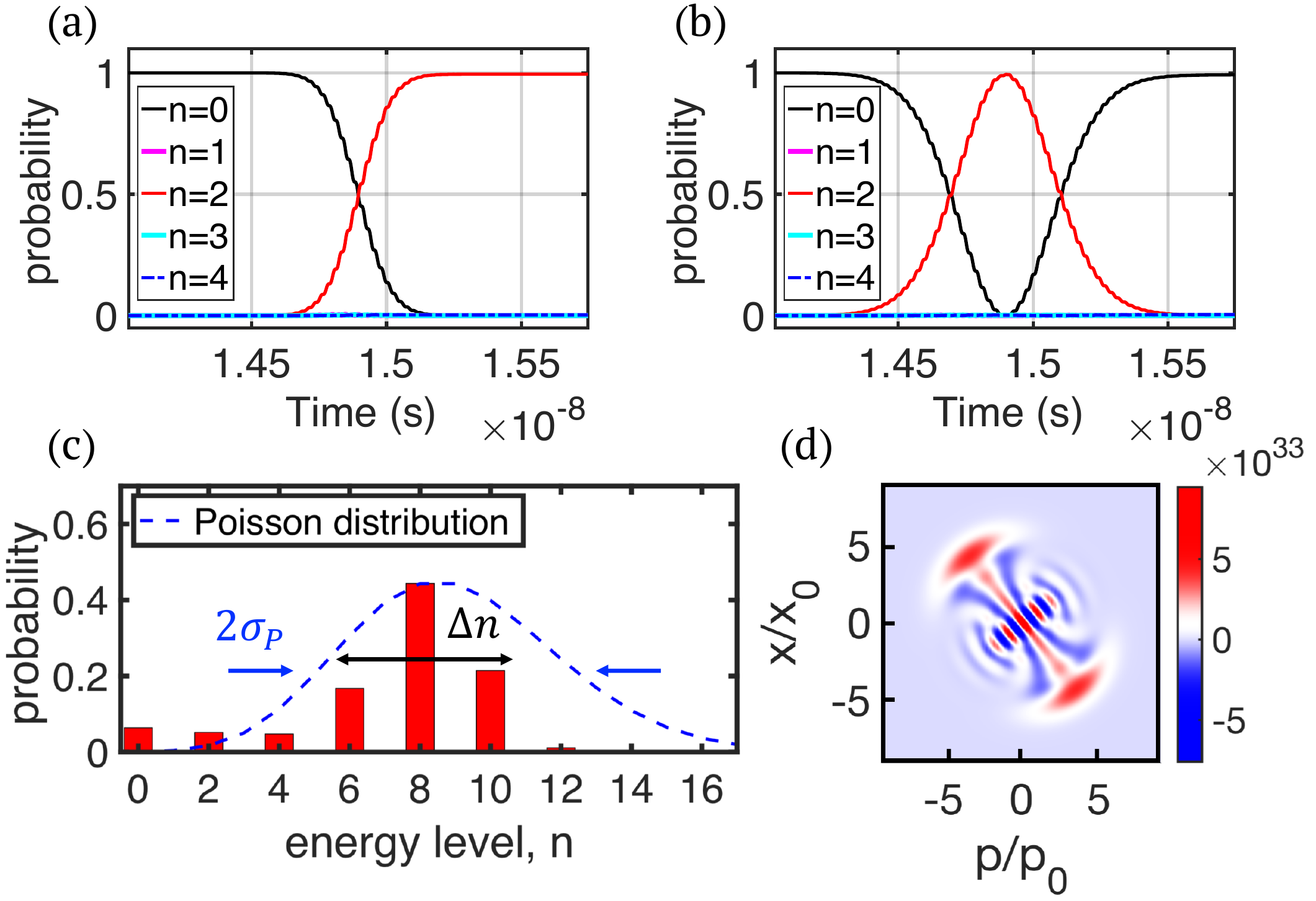}}
\caption{(color online) Kapitza-Dirac blockade as a tool to prepare energy eigenstates and Schr\"{o}dinger cat states of a harmonic oscillator. (a) Deterministic population transfer to the $\ket{n=2}$ eigenstate can be achieved with a pair of KD-laser pulses. Note that the $\ket{n=4}$ eigenstate (dashed line) is not populated as a result of the Kapitza-Dirac blockade. The small ripples on the probability trace are due to non-resonant excitation at $4\wo$. (b) When the pulse duration in (a) is doubled, the population is coherently returned to the ground state. (c) The population distribution of a Schr\"{o}dinger cat state has a sub-Poissonian width $\Delta n < 2\sigma_{_\textrm{P}}$, where $\sigma_{_\textrm{P}} = \sqrt{\nmax}$. (d) The Wigner function of the cat state indicates amplitude squeezing.}
\label{fig:electron}
\end{figure}

As a first example, we propose to prepare a single energy eigenstate $\ket{n = 2}$, starting from the ground state $\ket{n = 0}$. Setting $\deltap = 0.83$ suppresses the $\ket{n=2} \rightarrow \ket{n=4}$ transition down to $< 0.2\,\%$ of its maximum value (see \Fig{fig:transition_map}(c)). As a result, transitions starting in $\ket{n=0}$ will end deterministically in $\ket{n=2}$ without populating $\ket{n=4}$ or other excited states in the energy ladder (see \Fig{fig:electron}(a)). For a Gaussian-shaped time envelope, complete population inversion occurs when $\Omega_{R} \tkd  = \sqrt{2\pi}$, where $\Omega_{R}$ is the system specific Rabi frequency \cite{Suppl}. Note, that $\tkd$ should be sufficiently long to keep the pulse bandwidth below the trap frequency $\wo$ in order to avoid off-resonant excitation. When the pulse duration is doubled, the Rabi cycle is completed to the ground state (see \Fig{fig:electron}(b)). The Kapitza-Dirac blockade has thus transformed the harmonic oscillator into an effective two-level system --- with promising applications in quantum information processing. 

Kapitza-Dirac blockade can also prepare a Schr\"{o}dinger cat state. We show this with a heuristic example of $\deltap = -0.60$ which suppresses the transition $\ket{n=12} \rightarrow \ket{n=14}$ to $< 0.1\,\%$ of its maximum value (see \Fig{fig:transition_map}(b)). The oscillator then undergoes sequential excitation from $\ket{n=0}$ up to $\ket{n=12}$ in steps of $\Delta n = 2$, but $\ket{n=14}$ is not excited. The pulse intensity $\Ikd$ and duration $\tkd$ are adjusted together to maximize the population distribution at $\nmax = 8$ while avoiding off-resonant excitation. The width of the final population distribution is sub-Poissonian (see \Fig{fig:electron}(c)), due to the Kapitza-Dirac blockade. This leads to an amplitude-squeezed Schr\"{o}dinger cat state which is identified by inspecting the Wigner function in \Fig{fig:electron}(d). The maximum spatial and momentum separation of the cat state, $\Dx$ and $\Dp$, are
\begin{equation}\label{separation}
	\frac{\Dx}{\xo} = \frac{\Dp}{\hbar\ko} \approx 4\sqrt{\nmax}.
\end{equation}
Substituting \Eq{conservation} to \Eq{separation} with $\nnm=2$, the number of photon recoils is
\begin{equation}
	N_{\textrm{ph}} = \frac{\Dp}{\hbar(\kL+\kR)} \approx \frac{4\sqrt{\nmax}}{2+\deltap}, 
\end{equation}
which is again independent of the oscillator properties. If we take $\nmax = 650$ and $\deltap = -1.8$, the maximum momentum separation is $\Dp \approx 1000\, \hbar \klaser$, where $\klaser \equiv 2\pi/532\,\text{nm}$.

Kapitza-Dirac blockade is therefore a promising tool for realizing an all-optical large-momentum-transfer (LMT) beam splitter \cite{Kovachy2015, Gebbe2019, Chiow2011, Rudolph2020}. Moreover, the amplitude-squeezed Schr\"{o}dinger cat state leaves the divided beam rather well-collimated. A Kapitza-Dirac-LMT beam splitter used in conjunction with an optical Bragg grating \cite{Freimund2002, Giltner1995, Brand2020} could facilitate large-area matter-wave interferometry, in the future. 

\begin{figure}[t]
\centering
\scalebox{0.35}{\includegraphics{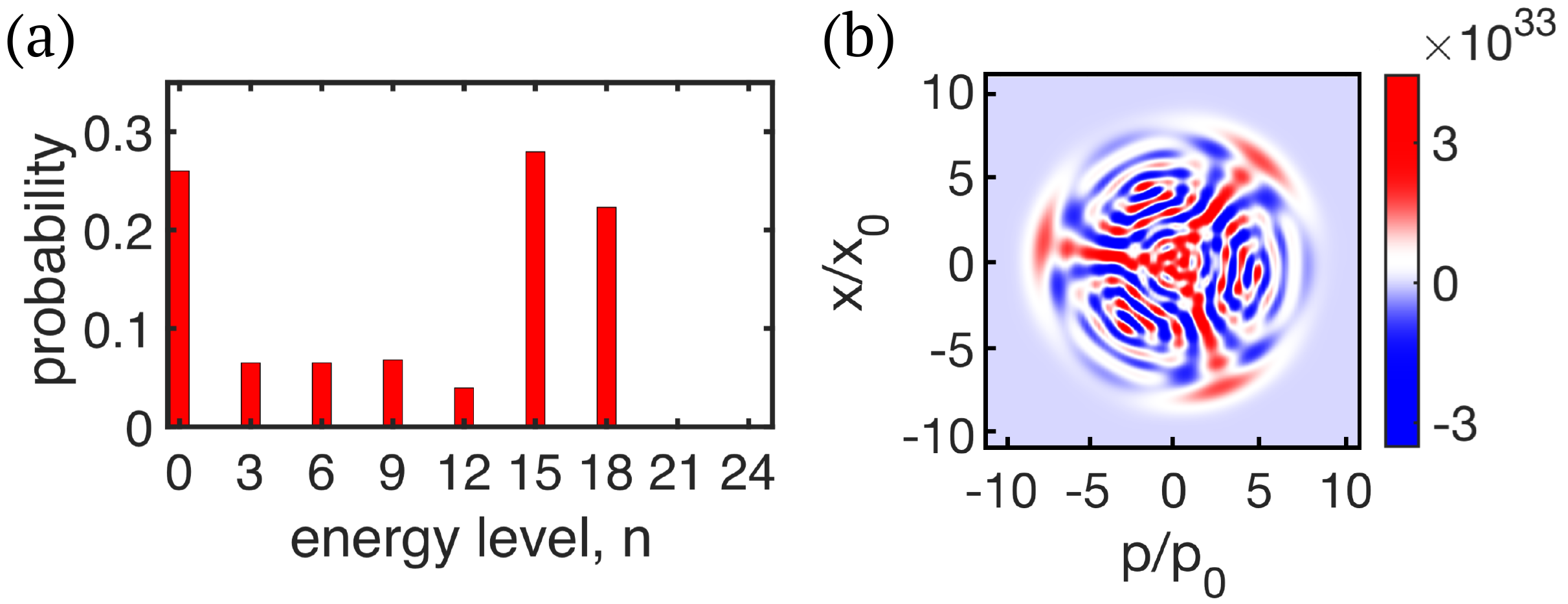}}
\caption{(color online) A 3-component Schr\"{o}dinger cat state. (a) A population distribution for a 3-component cat state can be prepared by taking $\nnm = 3$ and $\deltap = -1.57$ in \Eq{conservation}. (b) The Wigner function of the 3-component cat state shows interference fringes between each pair and among all three components.}.
\label{fig:cat3}
\end{figure}

Also multi-component cat states \cite{Vlastakis2013, Hofheinz2009, Johnson2017} can be prepared by properly choosing $\nnm$ and $\deltap$ \cite{Suppl}. An example is given in \Fig{fig:cat3} where a 3-component cat state is prepared. The KD-laser frequencies $\omega_{1,2}$ in this case are determined by taking $\nnm = 3$ and $\deltap = -1.57$, which suppresses the transition $\ket{n=18} \rightarrow \ket{n=21}$. As in multi-slit diffraction, interference among multiple components of the cat state produces sharper fringes compared to those between each pair (see \Fig{fig:cat3}(b)). Thus, a multi-component cat state can be more sensitive in probing quantum decoherence or collapse effects. 

Similarly, Gaussian states such as vacuum squeezed state ($\nnm = 2$) or coherent state ($\nnm = 1$) can be prepared with appropriate $\Ikd$ and $\deltap$.

Kapitza-Dirac blockade holds universally. Here we start the experimental discussion with the example of an electron beam 1D-trapped by the ponderomotive potential of a standing light wave \cite{Batelaan2007}. 
Close to the potential minimum, the potential can be approximated as a harmonic trap $U_{p}(x) \approx \left( \qe^2 \IS/2\epsilon_{0} c^3 \me \right) x^2$, where $\me$ and $\qe$ are the electron mass and charge, and $\IS$ is the standing wave intensity. The ponderomotive trap frequency is
\begin{equation}\label{ponderomotivetrap}
	\wo = \sqrt{\frac{\qe^2 \IS}{\epsilon_{0} c^3 \me^2}}.
\end{equation}
The intensity of each trapping laser that makes the standing wave is $\Itrap = \IS/4$. The trap ground state can be  populated by a well-collimated electron beam with transverse kinetic energy $mv^2_{x}/2 \ll \hbar\wo/2$.

If we consider the LMT beam splitter example above, using parameters as in \Table{table:par},  and if we tune the dynamics such that the electron reaches maximum momentum separation when it leaves the trap (see \Fig{fig:schematic}), it will leave in two distinct wave packets separated by 171\,$\mu$m in real space, already 1\,mm behind the trap \cite{Caprez2009, Hasselbach2010, Yasin2018}. 

All aspects discussed above can equally be realized with neutral massive particles: in \Table{table:par} we discuss as the second example a porphyrin derivative (TPPF84) with high mass and high vapor pressure \cite{Gerlich2011}. The third example is a silicon dioxide (SiO$_{2}$)  nanoparticle with low absorption of infrared trapping light.

The experimental schemes for all three systems can be  similar to that of \Fig{fig:schematic}. For molecules and nanoparticles a harmonic trap can be realized by the dipole potential of a standing wave. With $\Ltrap$ as the trapping laser wavelength, the dipole trap frequency is
\begin{equation}
	\Omega_{0} = \sqrt{ \frac{4\pi^2\alpha \IS}{\epsilon_{0} cm \Ltrap^2} },
\end{equation}
where $\alpha$ and $m$ are the particle's polarizability and mass. 

There are three criteria for choosing experimental parameters. First, the laser conditions should satisfy  $\Ttrap \gg \wzTL/2v_{z} \gg\wzKD/2v_{z} \gg \tkd \gg 40 (2\pi/\wo)$ \cite{Suppl}, where $\tau$ is the $1/e$ pulse duration, $W_{z}$ is the $1/e$ beam diameter along the $z$-axis, and $v_{z}$ is the particle speed in the $z$-direction. Second, the number of Rayleigh scattered photons should be small, $n_{sca}<1$, to avoid decoherence and dephasing. Third, the central KD-frequency $\wkd$ should be in the visible or the infrared regime because ultraviolet light would be absorbed in most materials. This implies that lower trap frequencies are preferred for more massive particles. The proposed experimental parameters for large amplitude-squeezed Schr\"{o}dinger cats of electrons, molecules, and nanoparticles are listed in \Table{table:par}, designed according to the empirical formulas
\begin{equation}
	\nbk \approx \frac{2}{3}\left( \frac{2\pi}{2+\deltap} \right)^2,
\end{equation}
\begin{equation}
	\Ikd\tkd \approx \frac{\mathcal{D}}{\mu_{0}}\sqrt{\frac{32\nbk\hbar m\wo}{\pi}},	
\end{equation}
where $\ket{\nbk}$ is the blockade state and the transition $\ket{\nbk} \rightarrow \ket{\nbk+2}$ is suppressed. The system specific coefficient $\mathcal{D}$ is $\me\wkd/\qe^2$ for electrons or $1/\alpha\wkd$ for molecules and nanoparticles. The momentum transfer $\Dp$ is well comparable with the state-of-the-art in matter-wave beam splitting \cite{Kovachy2015, Gebbe2019, Brand2020, Chiow2011, Rudolph2020}, but going beyond it as the concept can be applied universally to any 1D trapped particle that scatters light coherently. Additionally, the use of the Kapitza-Dirac blockade yields narrow momentum distributions and avoids overlap between the supposedly distinct wave packets. 

\begin{figure}[t]
\centering
\scalebox{0.55}{\includegraphics{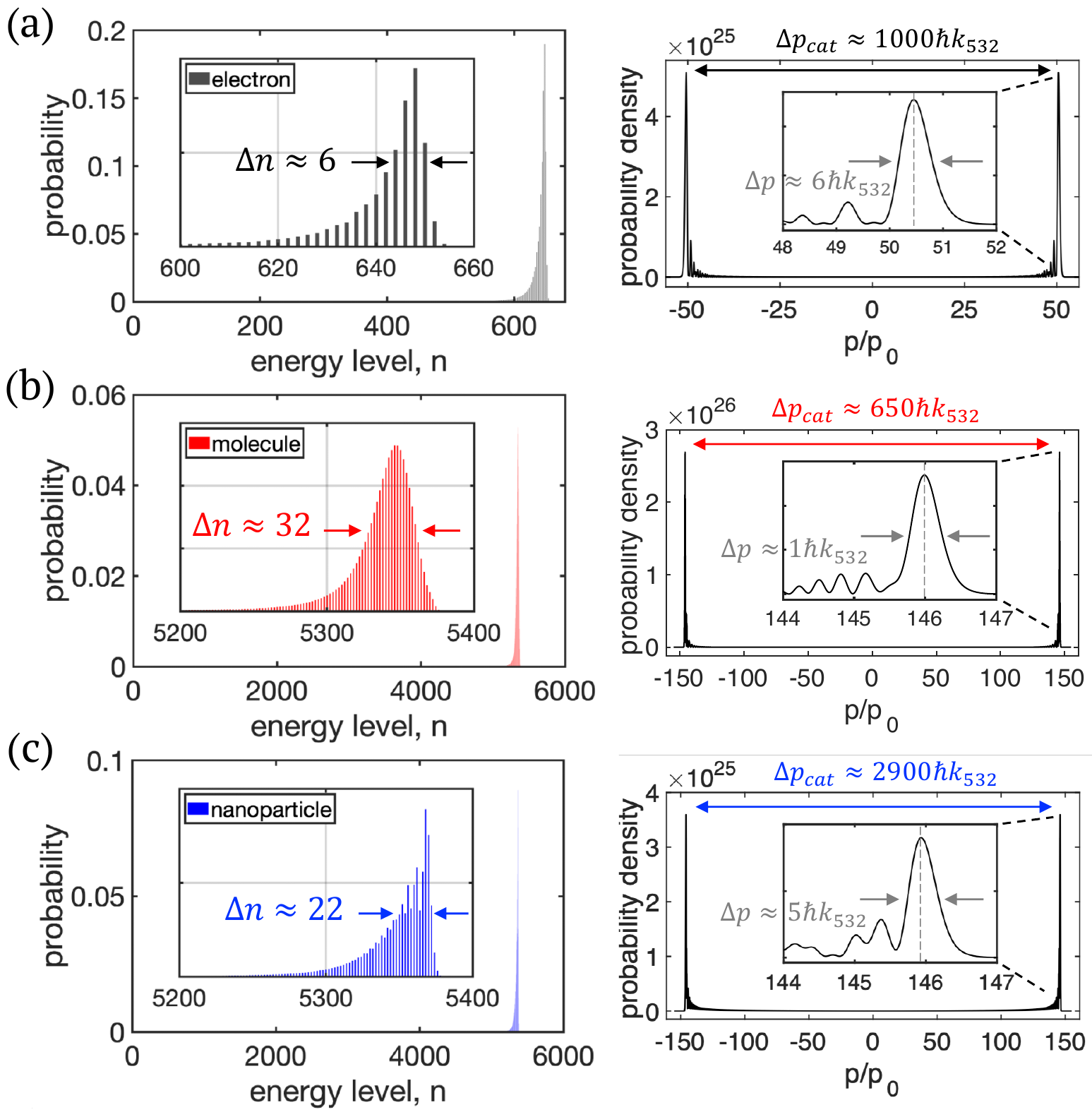}}
\caption{(color online) Simulated maximum momentum separation of amplitude-squeezed Schr\"{o}dinger cat states, for parameters as in \Table{table:par}. (a) electron: $\Dp \approx 1000 \hbar \klaser $. (b) molecule: $\Dp \approx 650 \hbar \klaser$. (c) nanoparticle: $\Dp \approx 2900 \hbar \klaser$. The population distributions for electron, molecule, and nanoparticle peak at $\nmax =$ 648, 5348, and 5368, respectively.}
\label{fig:cat2}
\end{figure}

The simulations were performed on a supercomputer using time-dependent Schr\"{o}dinger equations \cite{Suppl, Huang2015}. The results are shown in \Fig{fig:cat2}. The momentum transfer $\Dp$ is 1000 $\hbar \klaser$ for electrons, 650 $\hbar \klaser$ for molecules, and 2900 $\hbar \klaser$ for nanoparticles. The highest eigenstate available for excitation is determined by the onset of trap anharmonicity \cite{Suppl}. In the nanoparticle simulation, the maximal spatial separation in the trap is $\Dx \approx 1.2\,\mu m$. The required beam velocities for molecules and nanoparticles are two orders of magnitude lower than the state-of-the-art of free beams. However, cooling of molecules and nanoparticles is a rapidly advancing field and the proposed parameters are within reach \cite{Deachapunya2008, Patterson2015,Tarbutt2018,Asenbaum2013,Kuhn2017}.

In conclusion, we propose to use the Kapitza-Dirac blockade for manipulating the motional quantum states of single electrons, large molecules, and dielectric nanoparticles. The state preparation scheme is universal, applicable for different particles and independent of trap details and dimensions.

\begin{table}[t]
	\caption{Proposed parameters for preparing large amplitude-squeezed Schr\"{o}dinger cat states. The trapping laser for electrons has a $1/e$ pulse duration of $\Ttrap = 0.75$ ns and a repetition rate of 10 Hz. The trapping lasers for molecules and nanoparticles are continuous waves. The momentum detunings for electron, molecule (TPPF84), and nanoparticle (SiO$_{2}$) are $\deltap=$ -1.8, -1.93, and -1.93, respectively. $\langle P \rangle$ is the average power.} 
	\begin{ruledtabular}
		\begin{tabular}{cccc}
		parameters & electron & TPPF84 & SiO$_{2}$ nanoparticle	\\
		\hline
		$m$ (u) 							& $5.49 \times 10^{-4}$ & $2.81 \times 10^{3}$ & $10^{6}$  \\
		$\alpha$ (C$m^2$/V) 				& - & $2.22 \times 10^{-38}$ & $8.18 \times 10^{-36}$ 	\\
		$\wo$ (rad/s) 						& $3.22 \times 10^{12}$ & $3.13 \times 10^{4}$ & $1.75 \times 10^{3}$  \\
		$\Ltrap$ ($\mu$m) 					& 1.064 & 10.5 & 5	\\
		$\wyTL$ ($\mu$m) 					& 100 & 20 & 20 \\
		$\wzTL$ (mm)						& 1.6 & 5 & 9 	  \\
		$\langle P_{_\textrm{TL}} \rangle$ (W) 	& 37.6 & 30  & 0.037   \\
		$\IS (W/m^2)$						& $8 \times 10^{16}$ & $15 \times 10^{8}$ & $10.4 \times 10^{5}$ \\
		$v_{z}$ (m/s) 						& $6\times10^{6}$ & 0.1 & 0.02	 \\
		$\Lkd$ (nm) 						& 533 & 6819 & 1530	\\
		$\wyKD$ ($\mu$m) 					& 100 & 10 & 100 	\\
		$\wzKD$ (mm)						& 1.2 & 3 & 8	  \\
		$\langle P_{\textrm{KD}} \rangle$ (mW) 	& 400 & 280 & 1.5	 \\ 
		$\Ikd (W/m^2)$						& $2.6 \times 10^{15}$ & $1.2 \times 10^{7}$ & $2.3 \times 10^{3}$ \\
    		$\tkd$ (s) 							& $8 \times 10^{-11}$ & $1 \times 10^{-2}$ & $1.4 \times 10^{-1}$	\\
		$n_{sca}$							& $3.9 \times 10^{-3}$ & $8.8 \times 10^{-5}$  & $7.3 \times 10^{-1}$ 	 
		\end{tabular}
	\end{ruledtabular}
	\label{table:par}
\end{table}

Our simulations demonstrate the experimental feasibility of preparing various non-Gaussian states and large amplitude-squeezed Schr\"{o}dinger cat states. The latter has applications for all-optical LMT beam splitters in matter-wave interferometry. All these results explicitly rely on the coherent but destructive interference in the harmonic oscillator transition amplitudes in the presence of bichromatic light fields: the Kapitza-Dirac blockade. 

The proposed control scheme can also be employed for trapped ions and neutral atoms without invoking their internal states. In 2D or 3D harmonic traps one also finds entanglement among different motional degrees of freedom \cite{Suppl}. This can complement existing methods in quantum computing and quantum simulation.

\begin{acknowledgments}
The authors thank Aephraim M. Steinberg, Peter W. Milonni, Christopher Monroe, and Uro\v{s} Deli\'{c} for advice and discussions. W. C. Huang wishes to give a special thanks to Yanshuo Li for supports and helpful discussions. This work utilized high-performance computing resources from the Holland Computing Center of the University of Nebraska. Funding for this work comes from NSF EPS-1430519 and NSF PHY-1602755.
\end{acknowledgments}


\begin{thebibliography}{99}
	\bibitem{Brown1986} L. S. Brown and G. Gabrielse, Rev. Mod. Phys. \textbf{58}, 233 (1986).
	\bibitem{Hanneke2008} D. Hanneke, S. Fogwell, and G. Gabrielse, Phys. Rev. Lett. \textbf{100}, 120801 (2008).
	\bibitem{Leibried2003} D. Leibfried, R. Blatt, C. Monroe, D. Wineland, Rev. Mod. Phys. \textbf{75}, 281 (2003).
	\bibitem{Cornell2001} E. A. Cornell and C. E. Wieman, Rev. Mod. Phys. \textbf{74}, 875 (2001).
	\bibitem{Delic2020} U. Deli\'{c}, M. Reisenbauer, K. Dare, D. Grass, V. Vuleti\'{c}, N. Kiesel, M. Aspelmeyer, Science \textbf{367}, 892 (2020). 	
  	\bibitem{Arndt2014} M. Arndt and K. Hornberger, Nat. Phys. \textbf{10}, 271 (2014). 
	\bibitem{Zawisky2002}  M. Zawisky, M. Baron, R. Loidl, and H. Rauch, Nucl. Instrum. Methods Phys. Res. A \textbf{481}, 406 (2002).
	\bibitem{Kovachy2015} T. Kovachy, P. Asenbaum, C. Overstreet, C. A. Donnelly, S. M. Dickerson, A. Sugarbaker, J. M. Hogan, and M. A. Kasevich, Nature \textbf{528}, 530 (2015). 
	\bibitem{Gebbe2019} M. Gebbe, S. Abend, J. -N. Siem\ss, M. Gersemann, H. Ahlers, H. M\"{u}ntinga, S. Herrmann, N. Gaaloul, C. Schubert, K. Hammerer, C. L\"{a}mmerzahl, W. Ertmer, and E. M. Rasel, arXiv:1907.08416 (2019). 
	\bibitem{Fein2019} Y. Y. Fein, P. Geyer, P. Zwick, F. Kialka, S. Pedalino, M. Mayor, S. Gerlich, and M. Arndt, Nat. Phys. \textbf{15}, 1242 (2019).	
	\bibitem{Bateman2014} J. Bateman, S. Nimmrichter, K. Hornberger, and H. Ulbricht, Nat. Commun. \textbf{5}, 4788 (2014).  
	\bibitem{Bassi2013} A. Bassi, K. Lochan, S. Satin, T. P. Singh, and H. Ulbricht, Rev. Mod. Phys. \textbf{85}, 471 (2013).  
	\bibitem{Joos1985} E. Joos, and H. D. Zeh, Z. Phys. B. \textbf{59}, 223 (1985).
	\bibitem{Wootters1979} W. K. Wootters and W. H. Zurek, Phys. Rev. D \textbf{19}, 473 (1979). 
	\bibitem{Zurek1991} W. H. Zurek, Physics Today \textbf{44}, 36 (1991). 
	\bibitem{Bose2017} S. Bose, A. Mazumdar, G. W. Morley, H. Ulbricht, M. Toro\v{s}, M. Paternostro, A. A. Geraci, P. F. Barker, M. S. Kim, and G. Milburn, Phys. Rev. Lett. \textbf{119}, 240401 (2017). 
	\bibitem{Marletto2017} C. Marletto and V. Vedral, Phys. Rev. Lett. \textbf{119}, 240402 (2017). 
	\bibitem{Kapitza1933} P. L. Kapitza and P. A. M. Dirac, Math. Proc. Cambridge Philos. Soc. \textbf{29}, 297 (1933).  
	\bibitem{Freimund2001} D. L. Freimund, K. Aflatooni, and H. Batelaan, Nature \textbf{413}, 142 (2001).  
	\bibitem{Freimund2002} D. L. Freimund and H. Batelaan, Phys. Rev. Lett. \textbf{89}, 283602 (2002). 	
	\bibitem{Gould1986} P. L. Gould, G. A. Ruff, and D. E. Pritchard, Phys. Rev. Lett. \textbf{56}, 827 (1986). 	
	\bibitem{Martin1988} P. J. Martin, B. G. Oldaker, A. H. Miklich, and D. E. Pritchard, Phys. Rev. Lett. \textbf{60}, 515 (1988).  
	\bibitem{Pfau1994} T. Pfau, S. Sp\"{a}lter, Ch. Kurtsiefer, C. R. Ekstrom, and J. Mlynek, Phys. Rev. Lett. \textbf{73}, 1223 (1994). 
	\bibitem{Nairz2001} O. Nairz, B. Brezger, M. Arndt, and A. Zeilinger, Phys. Rev. Lett. \textbf{87}, 160401 (2001).  	
	\bibitem{Brand2020} C. Brand, F. Kia\l{}ka, S. Troyer, C. Knobloch, K. Simonovi\'{c}, B. A. Stickler, K. Hornberger, and M. Arndt, Phys. Rev. Lett. \textbf{125}, 033604 (2020).	
 	\bibitem{Talebi} N. Talebi and C. Lienau, New J. Phys. \textbf{21}, 093016 (2019).
	\bibitem{Schwartz} O. Schwartz, J. J. Axelrod, S. L. Campbell, C. Turnbaugh, R. M. Glaeser, and H. M\"{u}ller, Nat. Methods \textbf{16}, 1016 (2019).
	\bibitem{Huang2019} W. C. Huang and H. Batelaan, Atoms \textbf{7}, 42 (2019). 	
	\bibitem{Kasevich1991} M. Kasevich and S. Chu, Phys. Rev. Lett. \textbf{67}, 181 (1991).
	\bibitem{Kozuma1999} M. Kozuma, L. Deng, E. W. Hagley, J. Wen, R. Lutwak, K. Helmerson, S. L. Rolston, and W. D. Phillips, Phys. Rev. Lett. \textbf{82}, 871 (1999).
	\bibitem{Hemmerich1994} A. Hemmerich, C. Zimmermann, and T. W. H\"{a}nsch, Phys. Rev. Lett. \textbf{72}, 625 (1994).
	\bibitem{Monroe1995} C. Monroe, D. M. Meekhof, B. E. King, S. R. Jefferts, W. M. Itano, D. J. Wineland, and P. Gould, Phys. Rev. Lett. \textbf{75}, 4011 (1995).
	\bibitem{Suppl} Please see derivations and further discussions in the Supplementary Materiel. 
	\bibitem{Cahill1969} K. E. Cahill and R. J. Glauber, Phys. Rev. \textbf{177}, 1857 (1969). 		
	\bibitem{Wineland1998} D. J. Wineland, C. Monroe, W. M. Itano, B. E. King, D. Leibfried, D. M. Meekhof, C. Myatt, and C. Wood, Fortschr. Phys. \textbf{46}, 363 (1998).	  
	\bibitem{Chiow2011} S.-w. Chiow, T. Kovachy, H. -C. Chien, and M. A. Kasevich, Phys. Rev. Lett. \textbf{107}, 130403 (2011).
	\bibitem{Rudolph2020} J. Rudolph, T. Wilkason, M. Nantel, H. Swan, C. M. Holland, Y. Jiang, B. E. Garber, S. P. Carman, and Jason M. Hogan, Phys. Rev. Lett. \textbf{124}, 083604 (2020).
	\bibitem{Giltner1995} D. M. Giltner, R. W. McGowan, and S. A. Lee, Phys. Rev. Lett. \textbf{75}, 2638 (1995).
	\bibitem{Vlastakis2013} B. Vlastakis, G. Kirchmair, Z. Leghtas, S. E. Nigg, L. Frunzio, S. M. Girvin, M. Mirrahimi, M. H. Devoret, and R. J. Schoelkopf, Science \textbf{342}, 607 (2013). 	 
	\bibitem{Hofheinz2009} M. Hofheinz, H. Wang, M. Ansmann, R. C. Bialczak, E. Lucero, M. Neeley, A. D. O'Connell, D. Sank, J. Wenner, John M. Martinis, and A. N. Cleland, Nature \textbf{459}, 546 (2009).   
	\bibitem{Johnson2017} K.G. Johnson, J.D. Wong-Campos, B. Neyenhuis, J. Mizrahi, and C. Monroe, Nat. Commun. \textbf{8}, 697 (2017). 
	\bibitem{Batelaan2007} H. Batelaan, Rev. Mod. Phys. \textbf{79}, 929 (2007).  	
	\bibitem{Caprez2009} A. Caprez, R. Bach, S. McGregor, and H. Batelaan, J. Phys. B: At. Mol. Opt. Phys. \textbf{42}, 165503 (2009).
	\bibitem{Hasselbach2010} F. Hasselbach, Rep. Prog. Phys. \textbf{73}, 016101 (2010).
	\bibitem{Yasin2018} F. S. Yasin, K. Harada, D. Shindo, H. Shinada, B. J. McMorran, and T. Tanigaki, Appl. Phys. Lett. \textbf{113}, 233102 (2018).
	\bibitem{Gerlich2011} S. Gerlich, S. Eibenberger, M. Tomandl, S. Nimmrichter, K. Hornberger, P. J. Fagan, J. T\"{u}xen, M. Mayor, and M. Arndt, Nat. Comm. \textbf{2}, 263 (2011).
	\bibitem{Huang2015} W. C. Huang and H. Batelaan, Found. Phys. \textbf{45}, 333 (2015).		
	\bibitem{Deachapunya2008} S. Deachapunya, P. J. Fagan, A. G. Major, E. Reiger, H. Ritsch, A. Stefanov, H. Ulbricht, and M. Arndt, Eur. Phys. J. D \textbf{46} 307 (2008).
	\bibitem{Patterson2015} D. Patterson and J. M. Doyle, Phys. Chem. Chem. Phys. \textbf{17}, 5372 (2015).
	\bibitem{Tarbutt2018} M. R. Tarbutt, Contemp. Phys. \textbf{59}, 356 (2018). 
	\bibitem{Asenbaum2013} P. Asenbaum, S. Kuhn, S. Nimmrichter, U. Sezer, and M. Arndt, Nat. Commun. \textbf{4}, 2743 (2013).
	\bibitem{Kuhn2017} S. Kuhn, G. Wachter, F. Wieser, J. Millen, M. Schneider, J. Schalko, U. Schmid, M. Trupke, and M. Arndt, Appl. Phys. Lett. \textbf{111}, 253107 (2017). 

\end{thebibliography}
\end{document}


\title{Supplementary Material}

\author{Wayne~Cheng-Wei~Huang}
\affiliation{Department of Physics and Astronomy, Northwestern University, Evanston, Illinois 60208, USA}

\author{Herman~Batelaan}
\affiliation{Department of Physics and Astronomy, University of Nebraska-Lincoln, Lincoln, Nebraska 68588, USA}

\author{Markus~Arndt}
\affiliation{Faculty of Physics, University of Vienna, Boltzmanngasse 5, A-1090 Vienna, Austria}

\maketitle 

This supplementary material is aimed for providing a detailed account for the quantum mechanical treatment of the inelastic Kapitza-Dirac (KD) effect. Additionally, we extend the analysis to 2D and 3D harmonic oscillators and show that not only the Kapitza-Dirac blockade persists to work with added dimensionality, but the possibility of entanglement also arises. 

\subsection{I. Derivation of the dimensionless transition amplitude $g_{n}{(\eta)}$ in \textit{inelastic} Kapitza-Dirac effect}

\subsubsection{(a) An electron in a 1D ponderomotive trap}

\begin{figure}[b]
\centering
\scalebox{0.8}{\includegraphics{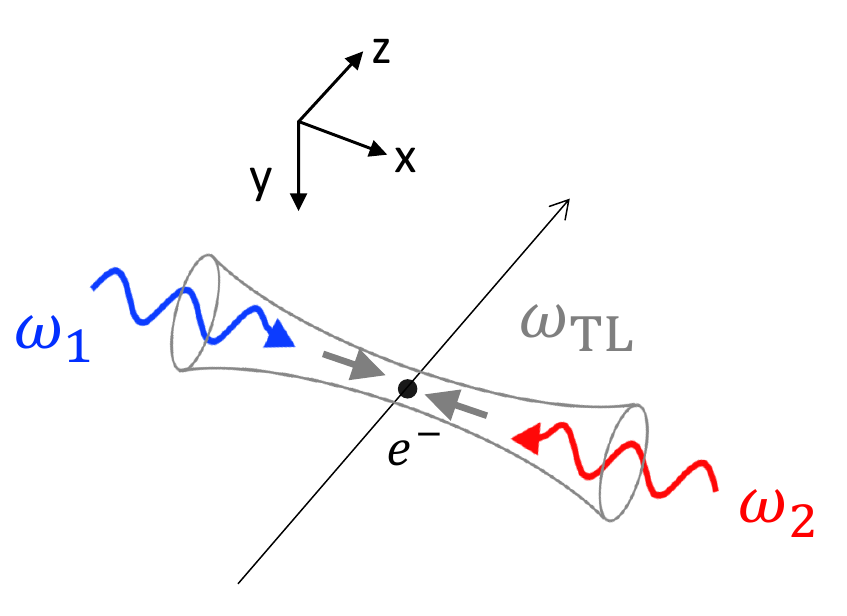}}
\caption{(color online) An electron interacting with a standing wave (gray) of frequency $\wS$ and two counter-propagating laser fields of frequencies $\omega_{1,2}$ (blue and red).}
\label{fig:electron}
\end{figure}

We consider a free electron interacting with two pairs of counter-propagating laser fields as shown in \Fig{fig:electron}. The electron travels in the $z$-direction, and the laser fields propagate in the $x$-direction. Both pairs of laser fields are pulsed. The first pair of fields has the same frequency $\wS$ and form a standing wave with a linear polarization along the $y$-axis. The second pair of fields are running waves with different frequencies $\omega_{1,2}$ ($\wL > \wR$) and a linear polarization along the $z$-axis. The quantum Hamilton of the electron is
\begin{equation}\label{hamiltonian0}
	\begin{split}
	 \hat{H} &= \frac{(\vect{p} - \qe\vect{A} )^2}{2\me} 	\\
	 	     &= \frac{\vect{p}^2}{2\me} - \frac{\qe}{2\me}\vect{p}\cdot\vect{A} - \frac{\qe}{2\me}\vect{A}\cdot\vect{p} + \frac{\qe^2}{2\me}\vect{A}^2, 	
	\end{split}
\end{equation} 
where $\vect{p} = -i\hbar\nabla$ and $\vect{A} = \AS + \AL + \AR$. We use $\qe = -e$ and $\me$ as electron charge and mass. The vector potential $\AS$ of the standing wave is
\begin{equation}\label{traplaser}
	\AS(x, t) = \sqrt{\frac{2\IS}{\epsilon_{0}c\wS^2}}\exp{(-t^2/\Ttrap^2)}\cos{(\ktrap x)}\cos{(\wS t)}\vec{\epsilon}_{y}, 
\end{equation}
where $\ktrap = \wS/c$ and $\IS$ is the intensity of the standing wave at the center of the waist. Note that the individual trapping laser beam in the standing wave has a pulse intensity $\Itrap = \IS/4$ and the $1/e$ pulse duration $\Ttrap$. We denote $x$ as the particle position operator hereafter. The pair of counter-propagating running waves are called KD-lasers, and their vector potential $\vect{A}_{1,2}$ are
\begin{equation}\label{KDlaser}
	\vect{A}_{1,2}(x, t) = A_{1,2}\exp{(-t^2/\tkd^2)}\cos{(- k_{1,2}x \pm \omega_{1,2}t)}\vec{{\epsilon}}_{z},
\end{equation}
where $A_{1,2} = \sqrt{2\Ikd/\epsilon_{0}c\omega_{1,2}^2}$, $k_{1,2} = \omega_{1,2}/c$, $\Ikd$ is the intensity of individual KD-lasers at the center of the waist, and $\tkd$ is the $1/e$ pulse duration. Here we approximate all laser fields to be homogenous over the area of interaction in the $y$-$z$ plane. This approximation is justified when (1) the laser beam sizes in the $y$-direction are much larger than the electron beam size along the same direction, and (2) the laser beam sizes in the $z$-direction satisfy the relation
\begin{equation}\label{beamsize}
	\Ttrap \gg  \frac{\wzTL}{2v_{z}} \gg \frac{\wzKD}{2v_{z}} \gg \tkd, 
\end{equation}
where $w_{z}$ is the $1/e$ beam diameter along the $z$-axis, and $v_{z}$ is the electron's speed along the $z$-axis. The subscripts $\textrm{-TL}$ and $\textrm{-KD}$ denote the trapping and the KD-lasers respectively.

The first term in \Eq{hamiltonian0} is the electron kinetic energy. The second term in \Eq{hamiltonian0} is simplified to be $\vect{p}\cdot\vect{A} = (-i\hbar\nabla)\cdot\vect{A} + \vect{A}\cdot(-i\hbar\nabla) = \vect{A}\cdot\vect{p}$ in the Coulomb gauge $\nabla\cdot\vect{A} = 0$. Thus it can be combined with the third term $\vect{A}\cdot\vect{p}$. At the first-order, the $\vect{A}\cdot\vect{p}$ term corresponds to single-photon scatterings which do not conserve energy and momentum for electrons. However, at the second order the $\vect{A}\cdot\vect{p}$ term can give rise to wave mixing and result in resonant scattering. From a classical analysis \cite{Huang2019}, it can be shown that the combined action of the $\vect{A}\cdot\vect{p}$ term and the $\vect{A}^2$ term in \Eq{hamiltonian0} is equivalent to keeping only the $\vect{A}^2$ term in \Eq{hamiltonian0}. Since the polarizations of the standing wave and the running waves are orthogonal $\AS\cdot \vect{A}_{1,2} = 0$, the $\vect{A}^2$ term can be expanded as $\vect{A}^2 =  \AS^2 + \AL^2 + \AR^2 + 2\AL\cdot\AR$ and the Hamiltonian in \Eq{hamiltonian0} can be rewritten as
\begin{equation}
	\hat{H} = \frac{\vect{p}^2}{2\me} + \frac{\qe^2}{2\me}( \AS^2 + \AL^2 + \AR^2 + 2\AL\cdot\AR).
\end{equation}
Assuming that the standing wave is much stronger than the running waves, $|\AS| \gg |\vect{A}_{1,2}|$, the Hamiltonian can be separated into an unperturbed part
\begin{equation}\label{unperturbedH}
	\hat{H}_{0} = \frac{\vect{p}^2}{2\me} + \frac{\qe^2\AS^2}{2\me} ,
\end{equation}
and an interaction part 
\begin{equation}\label{interactionH}
	\hat{H}_{int} =  \frac{\qe^2}{2\me} \left( \AL^2+\AR^2 + 2\AL\cdot\AR \right).
\end{equation}
Using \Eq{traplaser}, the unperturbed Hamiltonian can be written as $\hat{H}_{0} = \vect{p}^2/2\me + U(x,t)$, where
\begin{equation}\label{electrontrap0}
	U(x,t) = \frac{\qe^2\IS}{2\epsilon_{0}c\wS^2\me}\exp{(-2t^2/\Ttrap^2)}\cos^2{(\ktrap x)} \left( 1+ \cos{(2\wS t)} \right).
\end{equation}
The first term in \Eq{electrontrap0} is time-dependent, while the second term with $\cos{(2\wS t)}$ can be discarded because it only causes non-resonant driving and the effect averages out. Thus, we obtain the ponderomotive potential for the well-known elastic KD-effect,
\begin{equation}\label{electrontrap1}
	U_{p}(x) = \frac{\qe^2\IS}{2\epsilon_{0}c\wS^2\me}\cos^2{(\ktrap x)},
\end{equation}
where the Gaussian-shaped time envelope is approximated as $\exp{(-2t^2/\Ttrap^2)} \approx 1$ because it is slow-varying during the interaction time $\tkd$. Depending on the relation between the potential strength $U_{0} = \qe^2\IS/2\epsilon_{0}c\wS^2\me$, the recoil energy $E_{R} = \hbar^2(2\ktrap)^2/2\me$, and the transit time $\Delta t = \wzTL/v_{z}$, there are three parameter regimes of interaction: (1) diffraction regime ($E_{R} \ll \hbar/\Delta t$, $E_{R} \ll U_{0}$), (2) Bragg regime ($E_{R} \gg \hbar/\Delta t$, $E_{R} \gg U_{0}$), and (3) channeling regime ($U_{0} \gg E_{R} \gg \hbar/\Delta t$) \cite{Batelaan2000}. We assume that the experimental parameters are chosen for interaction in the channeling regime, so that the electron is transversely confined by the ponderomotive potential as it travels through the standing wave. Close to the potential minimum $x_{m} = \Ltrap/4$, where $\Ltrap = 2\pi/\ktrap$, the ponderomotive potential can be approximated as a harmonic trap
\begin{equation}
	U_{p}(x) \approx \frac{1}{2} \left( \frac{\qe^2\IS}{\epsilon_{0}c^3\me} \right) x^2 = \frac{1}{2}\me\wo^2 x^2,
\end{equation}
where $x$ is repurposed to denote the displacement from the potential minimum $x_{m}$, and the trap frequency is \begin{equation}
	\wo = \sqrt{\frac{\qe^2\IS}{\epsilon_{0}c^3\me^2}}.
\end{equation}
Therefore, the unperturbed Hamiltonian in \Eq{unperturbedH} becomes
\begin{equation}\label{H0-electron}
	\hat{H}_{0} = \left( \frac{p_{x}^2}{2\me}  + \frac{1}{2}\me\wo^2 x^2 \right) +  \frac{p_{y}^2+p_{z}^2}{2\me},
\end{equation}
where $(p_{x}, p_{y}, p_{z}) \equiv -i\hbar(\partial/\partial_{x}, \partial/\partial_{y}, \partial/\partial_{z})$. Thus, for parameters in the channeling regime the $x$-motion of the electron is confined in an 1D harmonic trap while the $y$- and $z$-motions freely propagate. The general solution to the electron's wavefunction is 
\begin{equation}\label{wavefunction0}
	\ket{\psi}(t) =  \phi(y, z, t) \left( \sum^{\infty}_{n=0} C_{n}(t)e^{-i\Omega_{n}t} \ket{n} \right), 
\end{equation} 
where $\phi(y, z, t)$ is the wavefunction of a free wavepacket, and $\Omega_{n} \equiv \wo(n + 1/2)$ characterizes the excitation of the harmonic oscillator in the $x$-direction. We are interested in solving the probability amplitude $C_{n}(t)$ for each eigenstate $\ket{n}$ of the harmonic oscillator. Using \Eq{KDlaser}, the interaction Hamiltonian in \Eq{interactionH} can be expanded as
\begin{equation}\label{hamiltonian1}
	\begin{split}
	\hat{H}_{int} =& \frac{\qe^2}{2\me} \exp{(-2t^2/\tkd^2)} \left[ A_{1}^2\cos^2{(\kL x - \wL t)} + A_{2}^2\cos^2{(\kR x + \wR t)} \right] 	\\
			&+ \frac{\qe^2}{2\me}A_{1}A_{2}  \exp{(-2t^2/\tkd^2)} \left[ \cos{\left((\kL+\kR)x - (\wL-\wR)t\right)} + \cos{\left((\kL-\kR)x - (\wL+\wR)t\right)} \right],
	\end{split}
\end{equation}
and the corresponding Schr\"{o}dinger equation for $C_{n}(t)$ is
\begin{equation}\label{Sequation1}
	i\hbar\frac{dC_{m}(t)}{dt} = \sum^{\infty}_{n=0} \bra{m} \hat{H}_{int} \ket{n}C_{n}(t)e^{i\Omega_{mn}t},
\end{equation}
where $\Omega_{mn} \equiv \Omega_{m} - \Omega_{n}$. To analyze the excitation dynamics, we group \Eq{Sequation1} into even and odd transitions,
\begin{equation}\label{Sequation2}
	\begin{split}
	i\hbar\frac{dC_{m}(t)}{dt} =& \sum^{\infty}_{n=0} \left(	\bra{m} \hat{H}_{int} \ket{m-2n}C_{m-2n}(t)e^{i2n\wo t}	 + \bra{m} \hat{H}_{int} \ket{m+2n}C_{m+2n}(t)e^{-i2n\wo t}	\right)	\\
						&+ \sum^{\infty}_{n=0} \left(\bra{m} \hat{H}_{int} \ket{m-(2n+1)}C_{m-(2n+1)}(t)e^{i(2n+1)\wo t} + \bra{m} \hat{H}_{int} \ket{m+(2n+1)}C_{m+(2n+1)}(t)e^{-i(2n+1)\wo t}	\right).
	\end{split}
\end{equation}
The terms in the first summation represent even transitions, while those in the second summation represent odd transitions. Assuming that $\omega_{1,2}$ are many orders of magnitude higher than $\wo$, only the $\wL-\wR$ term in \Eq{hamiltonian1} can resonantly excite the oscillator, which leads to a simplification of \Eq{hamiltonian1},
\begin{equation}\label{hamiltonian2}
	\begin{split}
	\hat{H}_{int}   =&   \frac{\qe^2}{4\me}A_{1}A_{2}  \exp{(-2t^2/\tkd^2)} \cos{\left((\kL+\kR)x\right)} \left( e^{i(\wL-\wR)t}+e^{-i(\wL-\wR)t} \right)	\\
			   & -i \frac{\qe^2}{4\me}A_{1}A_{2}  \exp{(-2t^2/\tkd^2)} \sin{\left((\kL+\kR)x\right)}\left( e^{i(\wL-\wR)t}-e^{-i(\wL-\wR)t} \right),
	\end{split}
\end{equation}
where $\cos{\left((\wL-\wR)t\right)}$ and $\sin{\left((\wL-\wR)t\right)}$ are expanded. In \Eq{hamiltonian2}, the first term with $\cos{\left((\kL+\kR)x\right)}$ gives rise to even transitions $\ket{n} \rightarrow \ket{n\pm2k}$, where $k$ is a positive integer. This is because the position operator of a harmonic oscillator can be expressed in terms of the rising and lowering operators, $x = \xo(\bd + \bb)$, where $\xo =\sqrt{\hbar/2\me\wo}$, $\bd\ket{n} = \sqrt{n+1} \ket{n+1}$, and $\bb\ket{n} = \sqrt{n} \ket{n-1}$. The expansion of $\cos{\left((\kL+\kR)x\right)}$ contains only even powers of $\bd$ and $\bb$. The resonant excitation frequency for even transitions $\ket{n} \rightarrow \ket{n\pm2k}$ is $\wL-\wR = 2k\wo$ because $e^{\pm i(\wL-\wR)t} = e^{\pm i2k\wo t}$ in \Eq{hamiltonian2} cancel with $e^{\mp i2n\wo t}$ in \Eq{Sequation2} for $n=k$. Similarly, the $\sin{\left((\kL+\kR)x\right)}$ term gives rise to the odd transitions, and the corresponding excitation frequency is $\wL-\wR = (2k+1)\wo$. 

The above observation can also be derived from the basis of parametric resonance. Let us switch to the interaction picture and replace the position operator $x$ in \Eq{hamiltonian2} by $x(t)$,
\begin{equation}
	x \longrightarrow x(t) = \sqrt{\frac{\hbar}{2\me\wo}}\left( \hat{b}^{\dagger}e^{i\wo t} + \hat{b}e^{-i\wo t} \right). 
\end{equation}
To obtain the frequency components of $\cos{\left((\kL+\kR)x\right)}$ and $\sin{\left((\kL+\kR)x\right)}$ in \Eq{hamiltonian2}, we first find the frequency components in $x^{m}(t)$ using the binomial expansion,
\begin{equation}
	x^{m}(t) = \left( \sqrt{\frac{\hbar}{2\me\wo}} \right)^{m} \sum^{m}_{n=0} e^{-i(m-2n)\wo t} {m \choose n} \hat{b}^{\dagger\,n} \bb^{(m-n)} .
\end{equation}
The frequency components, $|m-2n|\wo$, are either even or odd depending on the power $m$. Since there are only even powers of $x(t)$ in $\cos{\left((\kL+\kR)x\right)}$, the frequency components in $\cos{\left((\kL+\kR)x\right)}$ are all even. Similarly, the frequency components in $\sin{\left((\kL+\kR)x\right)}$ are all odd. Given the excitation frequency $\wL-\wR = 2k\wo$, there will be some frequency components in $\cos{\left((\kL+\kR)x\right)}$ that cancel with $e^{\pm i(\wL-\wR)t}$ and lead to parametric excitation. On the other hand, because $\cos{\left((\kL+\kR)x\right)}$ contains only even powers of $\bd$ and $\bb$, the transitions involved in parametric excitation at $\wL-\wR = 2k\wo$ are all even. 

A rigorous and general proof can be obtained by examining the transition matrix element of the interaction Hamiltonian in \Eq{hamiltonian2}, 
\begin{equation}\label{hamiltonian3}
	\bra{m} \hat{H}_{int}  \ket{n} = f(t) \bra{m}\cos{\left((\kL+\kR)x\right)}\ket{n}  + h(t) \bra{m} \sin{\left((\kL+\kR)x\right)}  \ket{n}, 
\end{equation}
where 
\begin{equation}\label{temporal}
	\begin{split}
		f(t) &\equiv (\qe^2/2\me) A_{1}A_{2}  \exp{(-2t^2/\tkd^2)} \cos{((\wL-\wR)t)}, 	\\
		h(t) &\equiv (\qe^2/2\me) A_{1}A_{2}  \exp{(-2t^2/\tkd^2)} \sin{((\wL-\wR)t)}.
	\end{split}
\end{equation}
We can find $\bra{m} \cos{\left((\kL+\kR)x\right)} \ket{n}$ and $\bra{m} \sin{\left((\kL+\kR)x\right)} \ket{n}$ by first evaluating the matrix element
\begin{equation}\label{generalKD1}
	\bra{m}e^{i(\kL+\kR)x}\ket{n} = \int^{\infty}_{-\infty} dx \phi_{m}^{\ast}(x)  e^{i(\kL+\kR)x} \phi_{n}(x),	
\end{equation}
where $\phi_{m}(x)$ is an eigen-wavefunction of the harmonic oscillator in real space. Using the convolution theorem 
\begin{equation}
	\int^{\infty}_{-\infty} dx\phi^{\ast}_{1}(x)\phi_{2}(x)e^{ik'x} = 2\pi\int^{\infty}_{-\infty} dk \psi^{\ast}_{1}(k)\psi_{2}(k-k'),
\end{equation}
where $\psi_{1,2}(k) = (1/2\pi) \int^{\infty}_{-\infty} dx \phi_{1,2}(x)e^{-ikx}$, \Eq{generalKD1} can be transformed to
\begin{equation}\label{generalKD2}
	\bra{m}e^{i(\kL+\kR)x}\ket{n} =  \int^{\infty}_{-\infty} dk \psi_{m}^{\ast}(k)\psi_{n}(k-(\kL+\kR)).
\end{equation}
Here we remark that \Eq{generalKD2} depicts the essential element in both elastic and inelastic KD-effect --- a periodic potential gives rise to momentum kicks. While \Eq{generalKD2} is general and valid for both free particles and harmonic oscillators, in the case of harmonic oscillators, there is one additional property due to the parity of the oscillator eigenstates $\ket{m}$ and $\ket{n}$,
\begin{equation}\label{generalKD3}
	 \int^{\infty}_{-\infty} dk \psi_{m}^{\ast}(k)\psi_{n}(k-(\kL+\kR)) = (-1)^{(m+n)} \int^{\infty}_{-\infty} dk \psi_{m}^{\ast}(k)\psi_{n}(k+(\kL+\kR)). 
\end{equation}
The physical interpretation of \Eq{generalKD3} is that for harmonic oscillators the transition amplitudes for stimulated emission (the integral with $\psi_{n}(k+(\kL+\kR))$) and absorption (the integral with $\psi_{n}(k-(\kL+\kR))$) are equal in strength but differ by a phase factor of the parity change $(-1)^{m+n}$. Using \Eqs{generalKD2}{generalKD3}, the transition matrix element in \Eq{hamiltonian3} can be evaluated as
\begin{equation}\label{hamiltonian4}
	\bra{m} \hat{H}_{int}  \ket{n} = \pi \left[ f(t) \left( 1+(-1)^{m+n} \right) -i h(t) \left( 1-(-1)^{m+n} \right) \right]\int^{\infty}_{-\infty} dk \psi_{m}^{\ast}(k)\psi_{n}(k-(\kL+\kR)).
\end{equation}
For even transitions $m = n+2k$, where $k$ is a positive integer, \Eq{hamiltonian4} becomes
\begin{equation}\label{matrixeven}
	\bra{n+2k} \hat{H}_{int}  \ket{n} = 2\pi f(t) \int^{\infty}_{-\infty} dk \psi_{n+2k}^{\ast}(k)\psi_{n}(k-(\kL+\kR))= f(t) \bra{n+2k}\cos{\left((\kL+\kR)x\right)}\ket{n}.
\end{equation}
For odd transition $m = n+2k+1$, \Eq{hamiltonian4} becomes
\begin{equation}\label{matrixodd}
	\bra{n+2k+1} \hat{H}_{int}  \ket{n} = -i2\pi h(t) \int^{\infty}_{-\infty} dk \psi_{n+2k+1}^{\ast}(k)\psi_{n}(k-(\kL+\kR))= h(t) \bra{n+2k+1}\sin{\left((\kL+\kR)x\right)}\ket{n}.
\end{equation}
If $\wL-\wR = 2k\wo$, it can be shown from \Eq{Sequation2} that only even transitions are resonantly driven. Their transition matrix elements can be evaluated by \Eq{matrixeven} which involves only the $\cos{\left((\kL+\kR)x\right)}$ term in \Eq{hamiltonian2}. Similarly, in the case of $\wL-\wR = (2k+1)\wo$ only the $\sin{\left((\kL+\kR)x\right)}$ term in \Eq{hamiltonian2} is involved in the resonant excitation. With \Eqs{Sequation2}{hamiltonian4}, we can summarize the conditions for resonant excitation in terms of energy-momentum conservation,
\begin{equation}\label{conservation}
	\left\{
	\begin{array}{l}
		\displaystyle	\wL - \wR = \nne \wo \\		\\
		\displaystyle	\kL + \kR = \left( \nne + \deltap \right) \ko	,
	\end{array}	
	\right.	
\end{equation}
where $\ko = \sqrt{\me\wo/2\hbar}$, $\nne$ is a positive integer, and $\deltap \ge -\nne$ is the momentum detuning discussed in the main text.

The energy-momentum relation in \Eq{conservation} and the transition amplitudes in \Eqs{matrixeven}{matrixodd} together give a clear physical picture of the inelastic KD-effect as an oscillator receiving a momentum kick $\hbar{(\kL+\kR)}$ and making a transition to a higher energy level. This picture allows us to give a physical interpretation for the Kapitza-Dirac blockade mechanism as described in FIG. 2 of the main text. However, the integral form of \Eqs{matrixeven}{matrixodd} are impractical to implement in numerical simulations, especially when the number of the involved eigenstates is large. Additionally, to identify the exact value of the momentum detuning $\deltap$ for Kapitza-Dirac blockade, it is desirable to derive the analytic forms for \Eqs{matrixeven}{matrixodd}. Cahill and Glauber \cite{Cahill1969} have shown that for a harmonic oscillator,
\begin{equation}\label{matrixelm}
	\bra{m}e^{i(\kL+\kR)x}\ket{n} = \sqrt{\frac{n!}{m!}}(i\eta)^{m-n}e^{-\eta^2/2}L^{(m-n)}_{n}(\eta^2)	\;\;\; \mathrm{for} 	\;\;\; m > n,
\end{equation}
where $\eta \equiv (\kL+\kR)\xo$ is the Lamb-Dicke parameter, and $L^{(m-n)}_{n}(y)$ is the generalized Laguerre polynomial. This equation gives \Eqs{matrixeven}{matrixodd} the desired analytic forms,
\begin{equation}\label{analytics}
	\begin{split}
	\bra{n+2k} \cos{\left((\kL+\kR)x\right)}  \ket{n}  &=  \sqrt{\frac{n!}{(n+2k)!}}(-1)^{k}\eta^{2k}e^{-\eta^2/2}L^{(2k)}_{n}(\eta^2),	\\
	\bra{n+2k+1} \sin{\left((\kL+\kR)x\right)}  \ket{n}  &=  \sqrt{\frac{n!}{(n+2k+1)!}}(-1)^{k}\eta^{2k+1}e^{-\eta^2/2}L^{(2k+1)}_{n}(\eta^2).
	\end{split}
\end{equation}

\begin{figure}[t]
\centering
\scalebox{0.7}{\includegraphics{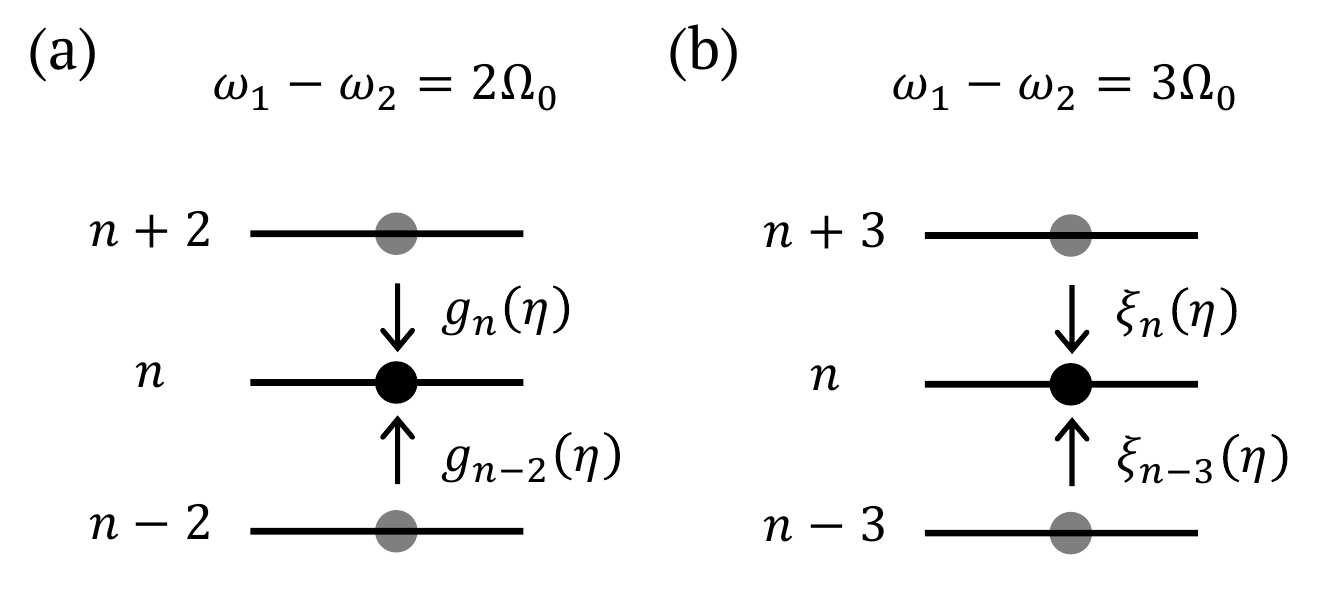}}
\caption{A schematic for the transition processes with (a) $\wL-\wR = 2\wo$ and (b) $\wL-\wR = 3\wo$.}
\label{fig:transition}
\end{figure}

So far we have been discussing the general theory of the inelastic KD-effect. Let us apply these results to the specific cases discussed in the main text. In the case of $\wL-\wR = 2\wo$, which is used in the main text to populate a single eigenstate $\ket{n=2}$ and to prepare a Schr\"{o}dinger cat state, the interaction Hamiltonian in \Eq{hamiltonian2} is reduced to \begin{equation}\label{hamiltonian5}
	\hat{H}_{int}   =  f(t) \cos{\left((\kL+\kR)x\right)} , 
\end{equation}
where $f(t)$ is given in \Eq{temporal}. Note that this interaction Hamiltonian is the effective KD-potential $\Vkd(\hat{x}, t)$ in the main text. Using \Eq{analytics}, we define the dimensionless transition amplitude
\begin{equation}\label{strength}
	g_{n}{(\eta)} \equiv \bra{n+2} \cos{\left((\kL+\kR)x \right)} \ket{n} = -\sqrt{\frac{n!}{(n+2)!}} \eta^{2} L^{(2)}_{n} (\eta^2) e^{-\eta^2/2}.
\end{equation}
Therefore, according to \Eq{Sequation2} the corresponding Schr\"{o}dinger equation is
\begin{equation}\label{Sequation3}
	i\hbar\frac{dC_{n}(t)}{dt} = f(t) \left( g_{n-2}{(\eta)}C_{n-2}(t)e^{i2\wo t} + g_{n}{(\eta)}C_{n+2}(t)e^{-i2\wo t} \right)	,
\end{equation}
where hermiticity is used to give $\bra{n+2} \hat{H}_{int} \ket{n} = \bra{n} \hat{H}_{int} \ket{n+2}$. We note that even in the Lamb-Dicke regime $\eta \ll 1$, the inelastic KD-effect as described in \Eq{Sequation3} can still be driven. For $\wL-\wR = 3\wo$, which is used in the main text to prepare a 3-component Schr\"{o}dinger cat state, the interaction Hamiltonian in \Eq{hamiltonian2} is reduced to 
\begin{equation}
	\hat{H}_{int}   =   h(t) \sin{\left((\kL+\kR)x\right)}.
\end{equation}
The dimensionless transition amplitude $\xi_{n}{(\eta)} \equiv \bra{n+3} \sin{\left((\kL+\kR)x \right)} \ket{n} = -\sqrt{n!/(n+3)!} \eta^{3} L^{(3)}_{n} (\eta^2) e^{-\eta^2/2}$ is used to identify the proper momentum detuing $\deltap = -1.57$ for the Kapitza-Dirac blockade at $\ket{n=18}$. The corresponding Schr\"{o}dinger equation is
\begin{equation}\label{Sequation4}
	i\hbar\frac{dC_{n}(t)}{dt} = h(t) \left( \xi_{n-3}{(\eta)}C_{n-3}(t)e^{i3\wo t} + \xi_{n}{(\eta)}C_{n+3}(t)e^{-i3\wo t} \right)	.
\end{equation}
 In \Fig{fig:transition}, we give an illustration for the transition processes in \Eqs{Sequation3}{Sequation4}.

\subsubsection{(b) A polarizable particle in a 1D dipole trap}
Next, we consider the interaction between two pairs of counter-propagating laser fields and a polarizable neutral particle (i.e. atoms, molecules, and dielectric nanoparticles). The proposed experimental setup resembles that in \Fig{fig:electron} with the only difference that the lasers are continuous-wave instead of pulsed. This is because the time scale of dynamics as determined by the trap frequency is much longer for massive particles compared to electrons. Polarizable neutral particles interact with laser fields through the dipole force, so the quantum Hamiltonian is 
\begin{equation}\label{Hpol0}
	\hat{H} = \frac{\vect{p^2}}{2M} - \frac{1}{2}\alpha\vect{E}^2,
\end{equation}
where $\alpha$ and $M$ are the  polarizability and mass of the particle, and $\vect{E} = \ES + \EL + \ER$.  For most large molecules and dielectric nanoparticles, the optical polarizability is approximately equal to the static polarizability. For atoms, the near-resonant enhancement in the optical polarizability can be used as a tuning parameter. Since the polarizations of the standing wave and the running waves are orthogonal $\ES\cdot \vect{E}_{1,2} = 0$, the $\vect{E}^2$ term can be expanded as $\vect{E}^2 =  \ES^2 + \EL^2 + \ER^2 + 2\EL\cdot\ER$ and the Hamiltonian in \Eq{Hpol0} can be rewritten as
\begin{equation}
	\hat{H} = \frac{\vect{p}^2}{2M} - \frac{1}{2}\alpha( \ES^2 + \EL^2 + \ER^2 + 2\EL\cdot\ER).
\end{equation}
The electric field of the trapping laser is 
\begin{equation}\label{Etrap0}
	\ES(x, t) = \sqrt{\frac{2\IS}{\epsilon_{0}c}}\cos{(\ktrap x)}\cos{(\wS t)}\vec{\epsilon}_{y}.
\end{equation}
The electric fields of the KD-lasers are 
\begin{equation}\label{KDfieldE}
	\vect{E}_{1,2}(x, t) = E_{1,2}\exp{(-t^2/\tkd^2)}\cos{(- k_{1,2}x \pm \omega_{1,2}t)}\vec{{\epsilon}}_{z},
\end{equation}
where $E_{1,2} = \sqrt{2\Ikd/\epsilon_{0}c}$. The KD-laser pulse duration $\tkd$ controls the interaction time, and it can be set by an acoustic optical modulator. As in \Eqs{traplaser}{KDlaser}, we approximate both the trapping and the KD-lasers to be homogeneous in the $y$-$z$ plane, assuming that the length scale of intensity variation is much larger than the particle beam size along the same direction. We also assume that the KD-laser fields satisfy the criteria given in \Eq{beamsize}. Assuming that the trapping laser is much stronger than the KD-lasers, $|\ES| \gg |\vect{E}_{1,2}|$, the Hamiltonian can be separated into an unperturbed part
\begin{equation}\label{Hpol4}
	\hat{H}_{0} =  \frac{\vect{p}^2}{2M} - \frac{1}{2}\alpha\ES^2, 
\end{equation}
and an interaction part
\begin{equation}\label{Hpol1}
	\hat{H}_{int} = - \frac{1}{2}\alpha(  \EL^2 + \ER^2 + 2\EL\cdot\ER).
\end{equation}
Using \Eq{Etrap0}, the unperturbed Hamiltonian can be written as $\hat{H}_{0} = \vect{p}^2/2M + U(x,t)$, where
\begin{equation}\label{moleculetrap0}
	U(x,t) = -\frac{\alpha\IS}{2\epsilon_{0}c}\cos^2{(\ktrap x)} \left( 1+ \cos{(2\wS t)} \right).
\end{equation}
The first term in \Eq{moleculetrap0} is time-independent. Assuming $\wS$ to be far from any electronic resonance of the particle, the second term with $\cos{(2\wS t)}$ can be discarded because it only causes non-resonant driving and the effect averages out. As a result, we obtain the dipole potential from \Eq{moleculetrap0},
\begin{equation}\label{moleculetrap1}
	U_{d}(x) = - \frac{\alpha\IS}{2\epsilon_{0} c}\cos^2{(\ktrap x)} . 
\end{equation}
Close to the potential minimum $x_{m}=0$, the dipole potential can be approximated as a harmonic trap
\begin{equation}\label{dipoletrap}
	U_{d}(x) \approx \frac{1}{2}\left( \frac{4\pi^2\alpha\IS}{\epsilon_{0} c\Ltrap^2} \right)x^2 = \frac{1}{2}M\wo^2x^2,
\end{equation}
where a constant in the expansion is dropped out, and the trap frequency is
\begin{equation}
	\wo = \sqrt{\frac{4\pi^2\alpha\IS}{\epsilon_{0} cM\Ltrap^2}}.
\end{equation}
Thus, the unperturbed Hamiltonian in \Eq{Hpol4} becomes
\begin{equation}\label{Hpol2}
	\hat{H}_{0} = \left( \frac{p_{x}^2}{2M}  + \frac{1}{2}M\wo^2 x^2 \right) +  \frac{p_{y}^2+p_{z}^2}{2M}.
\end{equation}
Since \Eq{Hpol2} is the same as \Eq{H0-electron}, the general solution to the particles' wavefunction is also described by \Eq{wavefunction0}. We can examine the interaction between the KD-lasers and the oscillator by evaluating \Eq{Hpol1},
\begin{equation}\label{Hpol3}
	\begin{split}
	\hat{H}_{int} =& - \frac{\alpha}{2} \exp{(-2t^2/\tkd^2)} \left[ E_{1}^2\cos^2{(\kL x - \wL t)} + E_{2}^2\cos^2{(\kR x + \wR t)} \right] 	\\
			& - \frac{\alpha}{2} E_{1}E_{2}  \exp{(-2t^2/\tkd^2)} \left[ \cos{\left((\kL+\kR)x - (\wL-\wR)t\right)} + \cos{\left((\kL-\kR)x - (\wL+\wR)t\right)} \right],
	\end{split}
\end{equation}
which has the same form as \Eq{hamiltonian1}. Comparing \Eqs{Hpol2}{Hpol3} to \Eqs{H0-electron}{hamiltonian1}, we see that the only difference is a direct replacement of coefficients,
\begin{equation}
	\begin{split}
		\me 	&\longrightarrow  M	\\
		A_{1,2} 	&\longrightarrow  E_{1,2}	\\
		\frac{\qe^2}{2\me}	&\longrightarrow  -\frac{\alpha}{2}.
	\end{split}
\end{equation}
Therefore, all the analysis and results from \Eq{Sequation1} to \Eq{Sequation4} apply for polarizable neutral particles.

\subsubsection{(c) Extension to 2D and 3D harmonic traps}
We now extend our discussion to particles in 2D and 3D harmonic traps, which are common for atoms, ions, molecules, and dielectric nanoparticles. The analysis presented here focuses on 3D harmonic oscillators but the results can be adapted for 2D harmonic oscillators. Let us consider a non-degenerate 3D harmonic oscillator with trap frequencies $(\wx,\wy,\wz)$ along the $x$-, $y$-, and $z$-axes. We extend the unperturbed Hamiltonian in \Eq{Hpol2} to 3D, 
\begin{equation}\label{HthreeD0}
	\hat{H}_{0} = \left( \frac{p_{x}^2}{2M}  + \frac{1}{2}M\wx^2 x^2 \right) +   \left( \frac{p_{y}^2}{2M}  + \frac{1}{2}M\wy^2 y^2 \right) +  \left( \frac{p_{z}^2}{2M}  + \frac{1}{2}M\wz^2 z^2 \right). 
\end{equation}
The oscillator is assumed to be initially in the ground state $\ket{0}_{x}\ket{0}_{y}\ket{0}_{z}$, where $\ket{n}_{x,y,z}$ are the eigenstates of the 3D harmonic oscillator along the $x$-, $y$-, and $z$-axes. We define these three degrees of freedoms as the $x$-, $y$-, and $z$-oscillators hereafter. If the KD-lasers propagate along the $x$-axis, only the $x$-oscillator can be resonantly excited. The general solution to the oscillator's wavefunction is 
\begin{equation}\label{wavefunction}
	\ket{\psi}(t) =  \ket{0}_{y} e^{-i\wy t/2} \ket{0}_{z} e^{-i\wz t/2} \left( \sum^{\infty}_{n=0} C_{n}(t)e^{-i\Omega_{n}t} \ket{n}_{x} \right), 
\end{equation} 
where $\Omega_{n} \equiv \wx(n + 1/2)$. Since the dynamics of $C_{n}(t)$ pertains to only the $x$-oscillator, the analysis from the previous subsection applies here.

\begin{figure}[t]
\centering
\scalebox{0.7}{\includegraphics{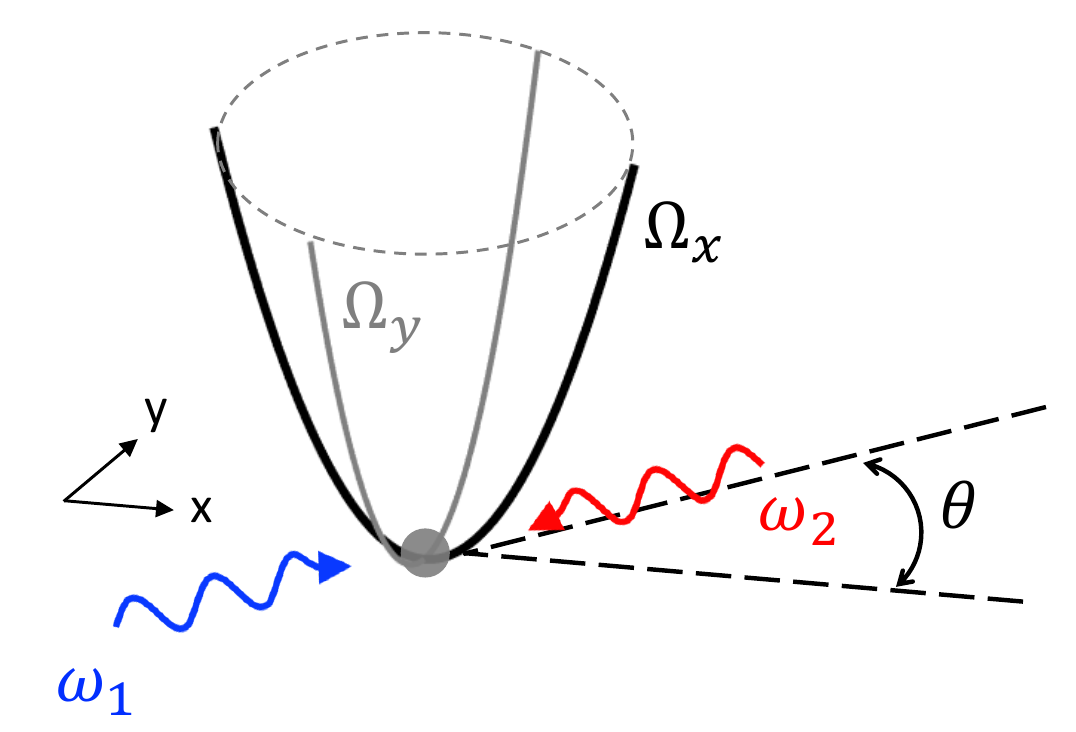}}
\caption{(color online) A 3D harmonic oscillator interacting with the KD-laser fields in the $xy$-plane.}
\label{fig:3Doscillator}
\end{figure}

\begin{figure}[t]
\centering
\scalebox{0.6}{\includegraphics{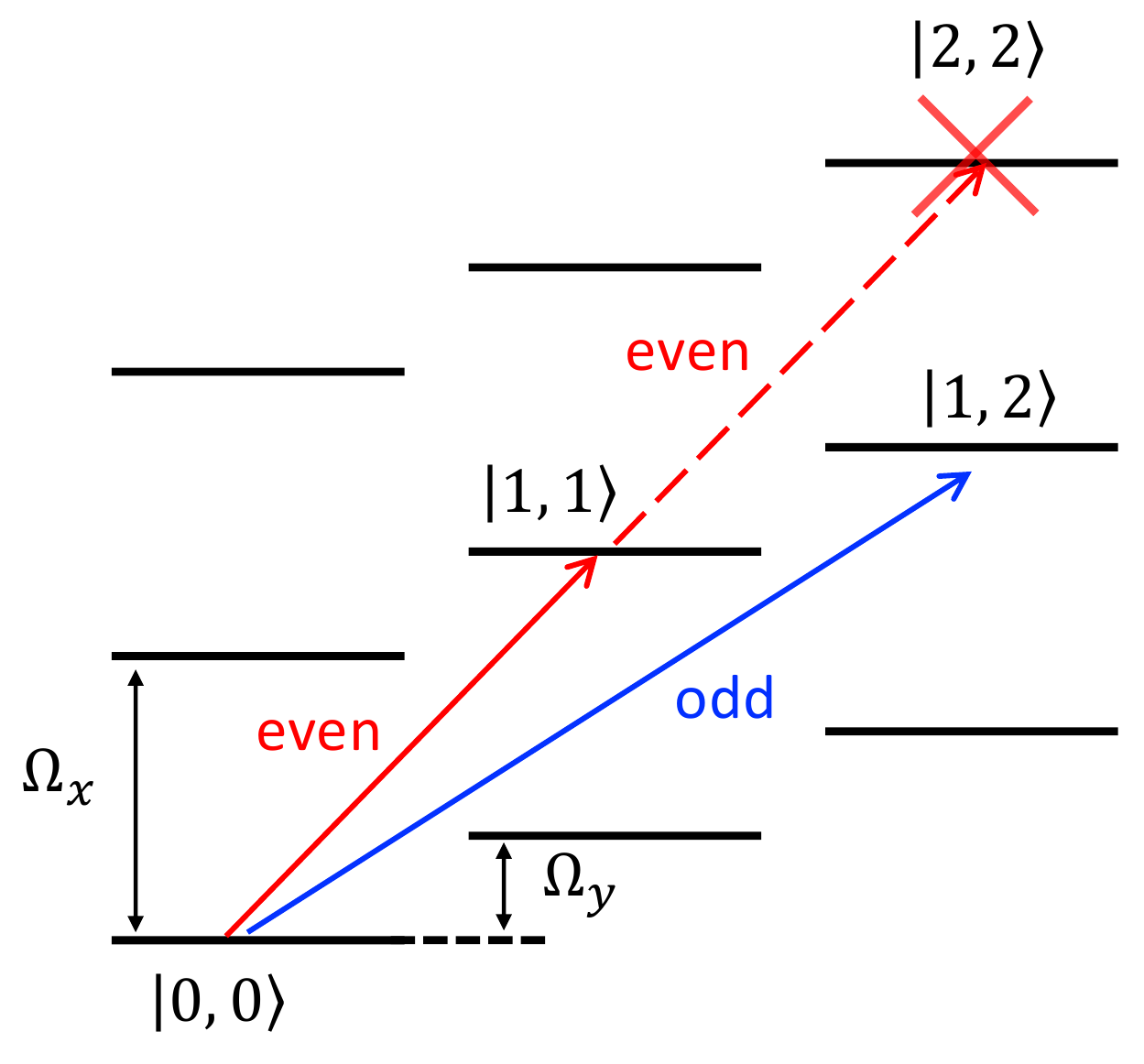}}
\caption{(color online) A partial level scheme of a 3D harmonic oscillator that undergoes even or odd transitions under inelastic Kapitza-Dirac scattering.}
\label{fig:level3D}
\end{figure}

When the KD-lasers travel in an oblique angle with respect to the potential axes, different degrees of freedom of the 3D harmonic oscillator can interact with the KD-laser fields at the same time and this leads to entanglement. As an example, we consider KD-laser fields traveling in the $xy$-plane with wavevectors $\vect{k}_{1,2} = \pm\left( \cos{(\theta)},\sin{(\theta)},0 \right)\omega_{1,2}/c$, where $\pi/2 \ge \theta \ge 0$ is the angle between the wavevector $\vect{k}_{1,2}$ and the $x$-axis. A schematic is given in \Fig{fig:3Doscillator}. The electric fields of the KD-lasers are modified from \Eq{KDfieldE} to be
\begin{equation}\label{3Dfield}
	\vect{E}_{1,2}(x, t) = E_{1,2}\exp{(-t^2/\tkd^2)}\cos{(- \vect{k}_{1,2}\cdot \vect{r} \pm \omega_{1,2}t)}\vec{{\epsilon}}_{z},
\end{equation}
where $\vect{r} = (x,y,z)$ is the 3D position operator of the particle. Because the interaction involve both the $x$- and $y$-oscillators, the general solution to the particle's wavefunction is modified from \Eq{wavefunction} to be
\begin{equation}
	\ket{\psi}(t) = \ket{0}_{z} e^{-i\wz t/2} \left( \sum^{\infty}_{m=0} \sum^{\infty}_{n=0} C_{m,n}(t)e^{-i \left( \Omega_{x}^{(m)}+\Omega_{y}^{(n)} \right) t} \ket{m,n} \right), 
\end{equation}
where $\ket{m,n} \equiv \ket{m}_{x}\ket{n}_{y} $, $\Omega_{x}^{(m)} \equiv \Omega_{x}(m + 1/2)$, and $\Omega_{y}^{(n)} \equiv \Omega_{y}(n + 1/2)$. The Schr\"{o}dinger equation for $C_{m,n}(t)$ is
\begin{equation}\label{Sequation3D}
	i\hbar\frac{dC_{m'n'}(t)}{dt} =  \sum^{\infty}_{m=0} \sum^{\infty}_{n=0}  \bra{m',n'} \hat{H}_{int} \ket{m,n}C_{m,n}(t)e^{i\left(\Omega_{x}^{(m')} - \Omega_{x}^{(m)}\right)t} e^{i\left(\Omega_{y}^{(n')} - \Omega_{y}^{(n)}\right)t}.
\end{equation}
Using \Eq{3Dfield}, the interaction Hamiltonian in \Eq{Hpol3} is modified to be
\begin{equation}\label{HthreeD5}
	\begin{split}
	\hat{H}_{int} =& - \frac{\alpha}{2} \exp{(-2t^2/\tkd^2)} \left[ E_{1}^2\cos^2{(\vect{k}_{1}\cdot \vect{r} - \wL t)} + E_{2}^2\cos^2{(\vect{k}_{2}\cdot \vect{r} + \wR t)} \right] 	\\
			& - \frac{\alpha}{2} E_{1}E_{2}  \exp{(-2t^2/\tkd^2)} \left[ \cos{\left(\kr  - (\wL-\wR)t\right)} + \cos{\left( \left(\vect{k}_{1}-\vect{k}_{2} \right)\cdot \vect{r} - (\wL+\wR)t\right)} \right].
	\end{split}
\end{equation}
Assuming $\omega_{1,2}$ to be many orders of magnitude higher than the trap frequencies $\Omega_{x,y,z}$, only the $\wL-\wR$ term in \Eq{HthreeD5} can resonantly excite the harmonic oscillator. Therefore, \Eq{HthreeD5} can be simplified as 
\begin{equation}\label{HthreeD6}
	\begin{split}
	\hat{H}_{int}   =&   - \frac{\alpha}{4}E_{1}E_{2}  \exp{(-2t^2/\tkd^2)} \cos{\left( \kr \right)} \left( e^{i(\wL-\wR)t}+e^{-i(\wL-\wR)t} \right)	\\
			   & + i \frac{\alpha}{4}E_{1}E_{2}  \exp{(-2t^2/\tkd^2)} \sin{\left( \kr \right)}\left( e^{i(\wL-\wR)t}-e^{-i(\wL-\wR)t} \right),
	\end{split}
\end{equation}
where $\cos{\left( (\wL-\wR)t \right)}$ and $\sin{\left( (\wL-\wR)t \right)}$ are expanded. The transition matrix element in \Eq{Sequation3D} is thus
\begin{equation}\label{HthreeD1}
	\bra{m',n'} \hat{H}_{int}  \ket{m,n} = f_{\alpha}(t) \bra{m',n'}\cos{\left( \kr \right)}\ket{m,n}  + h_{\alpha}(t) \bra{m',n'} \sin{\left( \kr \right)}  \ket{m,n}, 
\end{equation}
where 
\begin{equation}
	\begin{split}
		f_{\alpha}(t) &\equiv -(\alpha/2) E_{1}E_{2} \exp{(-2t^2/\tkd^2)} \cos{((\wL-\wR)t)}, 	\\
		h_{\alpha}(t) &\equiv -(\alpha/2) E_{1}E_{2} \exp{(-2t^2/\tkd^2)} \sin{((\wL-\wR)t)}.
	\end{split}
\end{equation}
Using \Eq{generalKD2}, we can evaluate the matrix element
\begin{equation}\label{matrix3D}
	\bra{m',n'} e^{i\kr} \ket{m,n} = \int^{\infty}_{-\infty} dk_{x} \psi_{m'}^{\ast}(k_{x})\psi_{m}(k_{x}-(\kL+\kR)_{x}) \int^{\infty}_{-\infty} dk_{y} \phi_{n'}^{\ast}(k_{y})\phi_{n}(k_{y}-(\kL+\kR)_{y}),
\end{equation}
where $\psi_{n}(k)$ and $\phi_{n}(k)$ are the eigen-wavefunctions of the $x$- and $y$-oscillators in the momentum space, and $(\kL+\kR)_{x,y} \equiv \vect{(\kL+\kR)}\cdot \vec{\epsilon}_{x,y}$ are the projection of $\vect{\kL+\kR}$ along the $x$- and $y$-axes. Using \Eq{matrix3D}, we transform \Eq{HthreeD1} into 
\begin{equation}\label{HthreeD2}
	\begin{split}
	\bra{m',n'} \hat{H}_{int}  \ket{m,n} =& \, \pi \left[ f_{\alpha}(t) \left( 1+(-1)^{m'+m+n'+n} \right) -i h_{\alpha}(t) \left( 1-(-1)^{m'+m+n'+n} \right) \right]	\\
		& \times \int^{\infty}_{-\infty} dk_{x} \psi_{m'}^{\ast}(k_{x})\psi_{m}(k_{x}-(\kL+\kR)_{x}) \int^{\infty}_{-\infty} dk_{y} \phi_{n'}^{\ast}(k_{y})\phi_{n}(k_{y}-(\kL+\kR)_{y}),
	\end{split}
\end{equation}
which is a generalization of \Eq{hamiltonian4}. With \Eqs{Sequation3D}{HthreeD2}, we can summarize the conditions for resonant excitation in terms of energy-momentum conservation,
\begin{equation}\label{conservation3D}
	\left\{
	\begin{array}{l}
		\displaystyle	\wL - \wR = \nnx \wx + \nny\wy  \\	\\
		\displaystyle	(\kL + \kR)_{x} = \left( \nnx + \delta_{px} \right) k_{x0}	\\	\\
		\displaystyle	(\kL + \kR)_{y} = \left( \nny + \delta_{py} \right) k_{y0}	,
	\end{array}	
	\right.	
\end{equation}
where $n_{x,y}$ are positive integers, $k_{x0} = \sqrt{M\wx/2\hbar}$, $k_{y0} = \sqrt{M\wy/2\hbar}$, and $\delta_{px, py}$ are the momentum detuning for the $x$- and the $y$-oscillator respectively. The central KD-laser frequency $\wkd$ and the laser propagation angle $\theta$ can be obtained from \Eq{conservation3D},
\begin{equation}
	\begin{split}
	\wkd &=  \frac{c}{2} \sqrt{\left( \nnx + \delta_{px} \right)^2 k_{x0}^2 + \left( \nny + \delta_{py} \right)^2 k_{y0}^2}	\\
	\theta &= \tan^{-1}{\left( \frac{\nny+\delta_{py}}{\nnx+\delta_{px}}\sqrt{\frac{\wy}{\wx}} \right)}, 
	\end{split}
\end{equation}
so the KD-laser frequencies are $\omega_{1,2} = \wkd \pm ( \nnx \wx + \nny\wy )/2$. We can see from \Eqs{HthreeD2}{conservation3D} that the joint parity change of the oscillator determines the symmetry of the transition, i.e. $(-1)^{m'+m+n'+n} = (-1)^{\nnx+\nny}$. If $\nnx$ and $\nny$ are both even or odd, then the transition is even and it involves only the $\cos{\left( \kr \right)}$ term in \Eq{HthreeD6}. Otherwise, the transition is odd. An example is provided in \Fig{fig:level3D}. Using \Eq{matrixelm}, we can evaluate \Eq{matrix3D} in terms of the generalized Laguerre polynomials,
\begin{equation}
	\begin{split}
	\bra{m',n'} e^{i\kr} \ket{m,n}  &= \bra{m'} e^{i(\kL+\kR)_{x}x} \ket{m} \bra{n'} e^{i(\kL+\kR)_{y}y} \ket{n}	  \\
						&= \left( \sqrt{\frac{m!}{m'!}}(i\eta_{x})^{m'-m}e^{-\eta_{x}^2/2}L^{(m'-m)}_{m}(\eta_{x}^2) \right) \left( \sqrt{\frac{n!}{n'!}}(i\eta_{y})^{n'-n}e^{-\eta_{y}^2/2}L^{(n'-n)}_{n}(\eta_{y}^2) \right)		
	\end{split}
\end{equation}
for $m' > m$ and $n' > n$. Note that the Lamb-Dicke parameters here are $\eta_{x} = (\nnx+\delta_{px})/2$ and $\eta_{y} = (\nny+\delta_{py})/2$. Assuming $m' = m + \nnx$ and $n' = n + \nny$, the matrix element in \Eq{HthreeD1} have the analytic forms 
\begin{equation}\label{matrixelement3D}
	\begin{split}
		\bra{m+\nnx,n+\nny}\cos{\left( \kr \right)}\ket{m,n} &= \frac{1}{2}\left( 1+(-1)^{\nnx+\nny} \right) (i)^{\nnx+\nny} F_{m,n}^{(\nnx,\nny)}(\eta_{x}, \eta_{y})	\\
		\bra{m+\nnx,n+\nny}\sin{\left( \kr \right)}\ket{m,n} &= \frac{1}{2i}\left( 1-(-1)^{\nnx+\nny} \right) (i)^{\nnx+\nny} F_{m,n}^{(\nnx,\nny)}(\eta_{x}, \eta_{y})	,
	\end{split}
\end{equation}
where
\begin{equation}
	F_{m,n}^{(\nnx,\nny)}(\eta_{x}, \eta_{y}) \equiv \sqrt{\frac{m!}{(m+\nnx)!}}\sqrt{\frac{n!}{(n+\nny)!}}(\eta_{x})^{\nnx}(\eta_{y})^{\nny}e^{-(\eta_{x}^2+\eta_{y}^2)/2}L^{(\nnx)}_{m}(\eta_{x}^2)L^{(\nny)}_{n}(\eta_{y}^2). 
\end{equation}

\noindent For $\wL - \wR = \nnx \wx + \nny\wy$, the Schr\"{o}dinger equation in \Eq{Sequation3D} can be simplified to include only the resonant terms, 
\begin{equation}\label{Sequation3Dsimple}
	\begin{split}
	i\hbar\frac{dC_{m,n}(t)}{dt} =&\, \bra{m,n} \hat{H}_{int} \ket{m-\nnx,n-\nny}C_{m-\nnx,n-\nny}(t)e^{i(\nnx\wx+\nny\wy) t} 	\\
						&+ \bra{m,n} \hat{H}_{int} \ket{m+\nnx,n+\nny}C_{m+\nnx,n+\nny}(t)e^{-i(\nnx\wx+\nny\wy) t}. 	
	\end{split}
\end{equation}
In the case of even transitions $\nnx+\nny = 2k$, where $k$ is a positive integer, the matrix element in \Eq{Sequation3Dsimple} can be evaluated from \Eq{matrixelement3D},
\begin{equation}\label{HthreeD3}
	\bra{m+\nnx,n+\nny} \hat{H}_{int}  \ket{m,n} = f_{\alpha}(t) \bra{m+\nnx,n+\nny}\cos{\left( \kr \right)}\ket{m,n} =  f_{\alpha}(t) (-1)^{k} F_{m,n}^{(\nnx,\nny)}(\eta_{x}, \eta_{y}).
\end{equation}
Similarly, in odd transitions $\nnx+\nny = 2k+1$, the matrix element in \Eq{Sequation3Dsimple} is
\begin{equation}\label{HthreeD4}
	\bra{m+\nnx,n+\nny} \hat{H}_{int}  \ket{m,n} = h_{\alpha}(t) \bra{m+\nnx,n+\nny}\sin{\left( \kr \right)}\ket{m,n} =  h_{\alpha}(t) (-1)^{k} F_{m,n}^{(\nnx,\nny)}(\eta_{x}, \eta_{y}).
\end{equation}
Let us examine the two examples of even and odd transitions in \Fig{fig:level3D}. For the even transition $(\nnx,\nny) = (1,1)$, \Eq{HthreeD3} becomes
\begin{equation}
	\bra{m+1,n+1} \hat{H}_{int}  \ket{m,n} = f_{\alpha}(t)g_{m,n}(\eta_{x},\eta_{y}) ,
\end{equation}
where
\begin{equation}\label{matrixeven3D}
	g_{m,n}(\eta_{x},\eta_{y}) \equiv  -F_{m,n}^{(1,1)}(\eta_{x}, \eta_{y}) = -\sqrt{\frac{m!}{(m+1)!}}\sqrt{\frac{n!}{(n+1)!}}(\eta_{x}\eta_{y})e^{-(\eta_{x}^2+\eta_{y}^2)/2}L^{(1)}_{m}(\eta_{x}^2)L^{(1)}_{n}(\eta_{y}^2).
\end{equation}
The corresponding Schr\"{o}dinger equation can be derived from \Eq{Sequation3Dsimple},
\begin{equation}
	i\hbar\frac{dC_{m,n}(t)}{dt} = f_{\alpha}(t) \left( g_{m-1,n-1}(\eta_{x},\eta_{y}) C_{m-1,n-1}(t)e^{i(\wx+\wy) t} + g_{m,n}(\eta_{x},\eta_{y}) C_{m+1,n+1}(t)e^{-i(\wx+\wy) t}  \right).
\end{equation}
For the odd transition $(\nnx,\nny) = (1,2)$, \Eq{HthreeD4} becomes
\begin{equation}
	\bra{m+1,n+2} \hat{H}_{int}  \ket{m,n} = h_{\alpha}(t)\xi_{m,n}(\eta_{x},\eta_{y}) ,
\end{equation}
where
\begin{equation}\label{matrixodd3D}
	\xi_{m,n}(\eta_{x},\eta_{y}) \equiv  -F_{m,n}^{(1,2)}(\eta_{x}, \eta_{y}) = -\sqrt{\frac{m!}{(m+1)!}}\sqrt{\frac{n!}{(n+2)!}}(\eta_{x}\eta_{y}^2)e^{-(\eta_{x}^2+\eta_{y}^2)/2}L^{(1)}_{m}(\eta_{x}^2)L^{(2)}_{n}(\eta_{y}^2),
\end{equation}
and the corresponding Schr\"{o}dinger equation is
\begin{equation}
	i\hbar\frac{dC_{m,n}(t)}{dt} = h_{\alpha}(t) \left( \xi_{m-1,n-2}(\eta_{x},\eta_{y}) C_{m-1,n-2}(t)e^{i(\wx+2\wy) t} + \xi_{m,n}(\eta_{x},\eta_{y}) C_{m+1,n+2}(t)e^{-i(\wx+2\wy) t}  \right).
\end{equation}
From \Eqs{matrixeven3D}{matrixodd3D}, we see that the sequential transition can be fully stopped at certain values of momentum detunings, $\delta_{px}$ or $\delta_{py}$. Therefore, Kapitza-Dirac blockade also works for 2D and 3D harmonic oscillators. Additionally, as the transition of one oscillator is suppressed, the transition of other oscillators stops too. Thus one set of laser parameters $(\wL, \wR, \theta)$ in \Eq{conservation3D} chosen for either $\delta_{px}$ or $\delta_{py}$ is sufficient for stopping transitions for both $x$- and $y$-oscillators. An entangled state $\ket{\psi} = (\ket{0,0} + \ket{1,1})/\sqrt{2}$ between the $x$- and $y$-oscillators can be prepared by suppressing the transition $\ket{1,1} \rightarrow \ket{2,2}$ as shown in \Fig{fig:level3D}. Complex entangled states such as a 2-component cat state entangling with a 3-component cat state (see FIG. 4(d) and FIG. 5 in the main text) can be prepared by using the odd transition $(\nnx,\nny) = (2,3)$ and placing the blockade on the transition $\ket{12,18} \rightarrow \ket{14, 21}$. Combination with a second pair of KD-lasers along the $y$-axis allows preparation of entangled states between cat states and single eigenstates, $\ket{\psi} = ((\ket{\alpha}_{x}+\ket{-\alpha}_{x})\ket{0}_{y} + (\ket{\alpha}_{x}-\ket{-\alpha}_{x})\ket{1}_{y})/2$, which make it possible to emulate a cavity-QED system with electrons, complex molecules, or dielectric nanoparticles.

\subsection{II. Wigner function and wave function in the discrete limit}

Given a wavefunction $\psi(x)$ in the configuration space, the Wigner function is \cite{Gerry2005} 
\begin{equation}\label{wignerfun}
	W(q, p) = \frac{1}{2\pi\hbar} \int_{-\infty}^{\infty} \psi^{\ast}\left( q-\frac{x}{2} \right) \psi\left( q+\frac{x}{2} \right) e^{ipx/\hbar} dx, 
\end{equation}
where $q$ and $p$ are position and momentum respectively. Defining $k \equiv p/\hbar$ and $f_{q}(x) \equiv \psi^{\ast}\left( q-x/2 \right) \psi\left( q+x/2 \right) $, we can rewrite \Eq{wignerfun} as a Fourier transform of $f_{q}(x)$,
\begin{equation}\label{wignerint}
	W(q,k) = \frac{1}{\hbar} \left( \frac{1}{2\pi}  \int_{-\infty}^{\infty} dx f_{q}(x)e^{ikx} \right). 
\end{equation}
As $(x,k)$ is a conjugated pair in Fourier transform, the integral can be discretized using the relation
\begin{equation}\label{integralconv}
	\frac{1}{2\pi} \int dx \leftrightarrow \frac{1}{L_{k}} \sum_{n} \Delta n , 
\end{equation}
where $\Delta n = 1$, and the discreteness of $x$ implies a finite size of the one-dimensional $k$-space $L_{k} = 2\pi/\Delta x$ with $\Delta x = x_{n+1} - x_{n}$. Using \Eq{integralconv}, we can transform \Eq{wignerint} into
\begin{equation}
	W(q,k) = \frac{1}{\hbar L_{k}} \sum_{n} f_{q}(x_{n}) e^{ikx_{n}}.
\end{equation}
A change of variable gives the discretized form of the Wigner function in \Eq{wignerfun},
\begin{equation}\label{wignerfun-disc}
	W(q,p) = \frac{\Delta x}{2\pi} \sum_{n} \psi^{\ast}\left( q-x_{n}/2 \right) \psi\left( q+x_{n}/2 \right) e^{ipx_{n}/\hbar}.
\end{equation}
The Wigner functions presented in FIG. 4 and FIG. 5 of the main text are computed through \Eq{wignerfun-disc}. Additionally, we note that the probability distributions $|\psi(x)|^2$ and $|\varphi(p)|^2$ in the position and momentum space can be obtained as marginals of the Wigner function,
\begin{equation}
	\begin{split}
	|\psi(x)|^2 &= \int_{-\infty}^{\infty} dp W(x,p) \leftrightarrow |\psi(x)|^2 = \frac{2\pi\hbar}{L_{x}} \sum_{n} W(x,p_{n}),	\\
	|\varphi(p)|^2 &= \int_{-\infty}^{\infty} dx W(x,p) \leftrightarrow |\varphi(p)|^2 = \frac{2\pi\hbar}{L_{p}} \sum_{n} W(x_{n},p),
	\end{split}
\end{equation}
where $L_{p} = \hbar L_{k}$.

In our numerical simulation of inelastic KD-effect, the time-dependent Schr\"{o}dinger equations in \Eq{Sequation3} and \Eq{Sequation4} are solved for a $N$-state harmonic oscillator. The real and imaginary part of the Schr\"{o}dinger equation are separated, so there are $2N$ real-number differential equations to be solved simultaneously. We employ the adaptive 5th order Cash-Karp Runge-Kutta method as the equation solver in our FORTRAN code \cite{Press1996}. The oscillator is assumed to be in the ground state initially. The integration time is 5 times longer than the $1/e$ pulse duration of the KD-lasers. The integration stepsize is $1/100$ of the natural period $2\pi/\wo$. The most computationally intense simulation is for a nanoparticle oscillator with $N = 6000$ eigenstates (see FIG. 6(c) in the main text). The simulation took 400 GB memory and 90 hours of computation time on a single core of a supercomputer (Crane, Holland Computing Center). Once we obtain the probability amplitude $C_{n}(t)$ for each eigenstate, we compute the oscillator's wavefunction using MATLAB, and the wavefunction is subsequently used for computing the Wigner function in \Eq{wignerfun-disc}. We note that for states $n > 170$ the  value of the normalization factor $\sqrt{1/2^n n!}$ in the analytical formula of the oscillator eigen-wavefunction is too large for the computer to compute. To circumvent this problem, we solve the eigen-wavefunction numerically using the time-independent Schr\"{o}dinger equation in the range of $0 \le x \le x_{t}$, where $x_{t} = \sqrt{(n+1/2)2\hbar/M\wo}$ is the turning point. In the tunneling regime $x \ge x_{t}$, we patch an ansatz function to the eigen-wavefunction 
\begin{equation}
	\phi(x \ge x_{t}) = \phi(x_{t})\exp{\left[ \frac{(x-x_{t})\phi'(x_{t})}{\phi(x_{t})} \right]}\exp{\left[-\frac{(x-x_{t})^2}{2\xo^2} \right]},
\end{equation}
where $\phi'(x) \equiv d\phi(x)/dx$ and $\xo = \sqrt{\hbar/2M\wo}$. The eigen-wavefunction in the range of $x < 0$ is obtained through symmetry $\phi(x < 0) = (-1)^n\phi(-x)$, where $n$ is the quantum number of the eigenstate. The numerically solved eigen-wavefunctions are in very good agreement with the analytic formula. The data size of the nanoparticle oscillator eigen-wavefunctions is 24 GB.

\subsection{III. Evaluation of the maximum spatial and momentum separation of a cat state}

We first consider a coherent state $\ket{\alpha}$(t) oscillating back and forth in a harmonic potential. The Hamiltonian for a harmonic oscillator with a mass $M$ and a natural frequency $\wo$ is
\begin{equation}
	\hat{H} = \frac{\pp^2}{2M} + \frac{1}{2}M\wo^2\xx^2 = \frac{(\pp-\pc)^2+2(\pp-\pc)\pc+\pc^2}{2M} +  \frac{1}{2}M\wo^2\left[ (\xx-\xc)^2+2(\xx-\xc)\xc+\xc^2 \right],
\end{equation}
where $\pc = \avg{\pp}(t)$ and $\xc = \avg{\xx}(t)$ are the momentum and the position expectation values of the wavepacket at time $t$. The energy expectation value is
\begin{equation}\label{energycat0}
	\avg{\hat{H}}(t) = \frac{\avg{(\pp-\pc)^2}+2\avg{\pp-\pc}\pc+\pc^2}{2M} +  \frac{1}{2}M\wo^2\left[ \avg{(\xx-\xc)^2}+2\avg{\xx-\xc}\xc+\xc^2  \right]. 
\end{equation}	
Because the Gaussian probability distribution of the coherent state is symmetric with respect to its average momentum and position, $\pc$ and $\xc$, we have $\avg{\pp-\pc} = 0$ and $\avg{\xx-\xc} = 0$, and \Eq{energycat0} is simplified to
\begin{equation}\label{energycat1}
	\avg{\hat{H}}(t) = \left( \frac{\avg{(\pp-\pc)^2}}{2M} +  \frac{1}{2}M\wo^2\avg{(\xx-\xc)^2} \right) + \left( \frac{\pc^2}{2M} + \frac{1}{2}M\wo^2\xc^2 \right).
\end{equation}
The terms in the first parenthesis are the kinetic and potential energies associated with the spread of the wavepacket, $\sigma_{x}^2 = \avg{(\xx-\xc)^2}$ and $\sigma_{p}^2 = \avg{(\pp-\pc)^2}$. The terms in the second parenthesis give the kinetic and potential energies associated with the average momentum and position of the wavepacket, which evolve as those of a classical harmonic oscillator according to the Ehrenfest theorem. Since the coherent state have the same Gaussian probability distribution as the ground state, the terms in the first parenthesis are equal to $\hbar\wo/2$, and \Eq{energycat1} is further simplified as
\begin{equation}\label{energycat2}
	\avg{\hat{H}}(t) = \frac{\hbar\wo}{2} + \left( \frac{\pc^2}{2M} + \frac{1}{2}M\wo^2\xc^2 \right).
\end{equation}

Now, we consider a cat state $\ket{\psi}_{cat}(t) = (\ket{\alpha}(t)+\ket{-\alpha}(t))/\sqrt{2}$ in the harmonic potential. We assume that the width of the two wavepackets $\ket{\alpha}$ and $\ket{-\alpha}$ are smaller than the maximum spatial separation $\Dx$ between the peaks of the two wavepackets. At the turning points, i.e. $\xc = \Dx/2$, the energy expectation values associated with each wavepackets can be calculated using \Eq{energycat2},
\begin{equation}\label{potentialE}
	\avg{E}_{1,2} = \frac{\hbar\wo}{2} + \frac{1}{2}M\wo^2\left( \frac{\Dx}{2}\right)^2.
\end{equation}
Meanwhile, the energy expectation value of the cat state can be calculated from the population distribution,
\begin{equation}\label{energycat3}
	\avg{E}_{cat} = \hbar\wo \left( \avg{n} + \frac{1}{2} \right).
\end{equation}
Assuming sufficiently large $|\alpha|$, i.e. well-separated wavepackets, we have $\langle -\alpha \ket{\alpha} = 0$, so $\avg{E}_{cat}$ can be related to $\avg{E}_{1,2}$ by
\begin{equation}\label{energycat4}
	\avg{E}_{cat} = \frac{1}{2} \left( \bra{\alpha}\hat{H}\ket{\alpha} + \bra{-\alpha}\hat{H}\ket{-\alpha}\right) = \frac{1}{2}(\avg{E}_{1} + \avg{E}_{2}).
\end{equation}
Substituting \Eqs{potentialE}{energycat3} to \Eq{energycat4}, we obtain a relation between the maximum spatial separation $\Dx$ and the quantum number at the peak of the population distribution $\nmax \equiv \avg{n}$,
\begin{equation}
	\frac{\Dx}{\xo} = 4\sqrt{\nmax}, 
\end{equation}
where $\xo = \sqrt{\hbar/2M\wo}$. Similarly, at the potential minimum, \Eq{energycat2} becomes
\begin{equation}\label{kineticE}
	\langle E \rangle_{1,2} = \frac{\hbar\wo}{2} + \frac{1}{2M}\left( \frac{\Dp}{2}\right)^2.
\end{equation}
Again, substituting \Eqs{energycat3}{kineticE} to \Eq{energycat4} we obtain a relation between the maximum momentum separation $\Dp$ and $\nmax$,
\begin{equation}
	\frac{\Dp}{\hbar\ko} = 4\sqrt{\nmax}, 
\end{equation}
where $\ko = \sqrt{ M\wo/2\hbar}$. 

\begin{figure}[t]
\centering
\scalebox{0.7}{\includegraphics{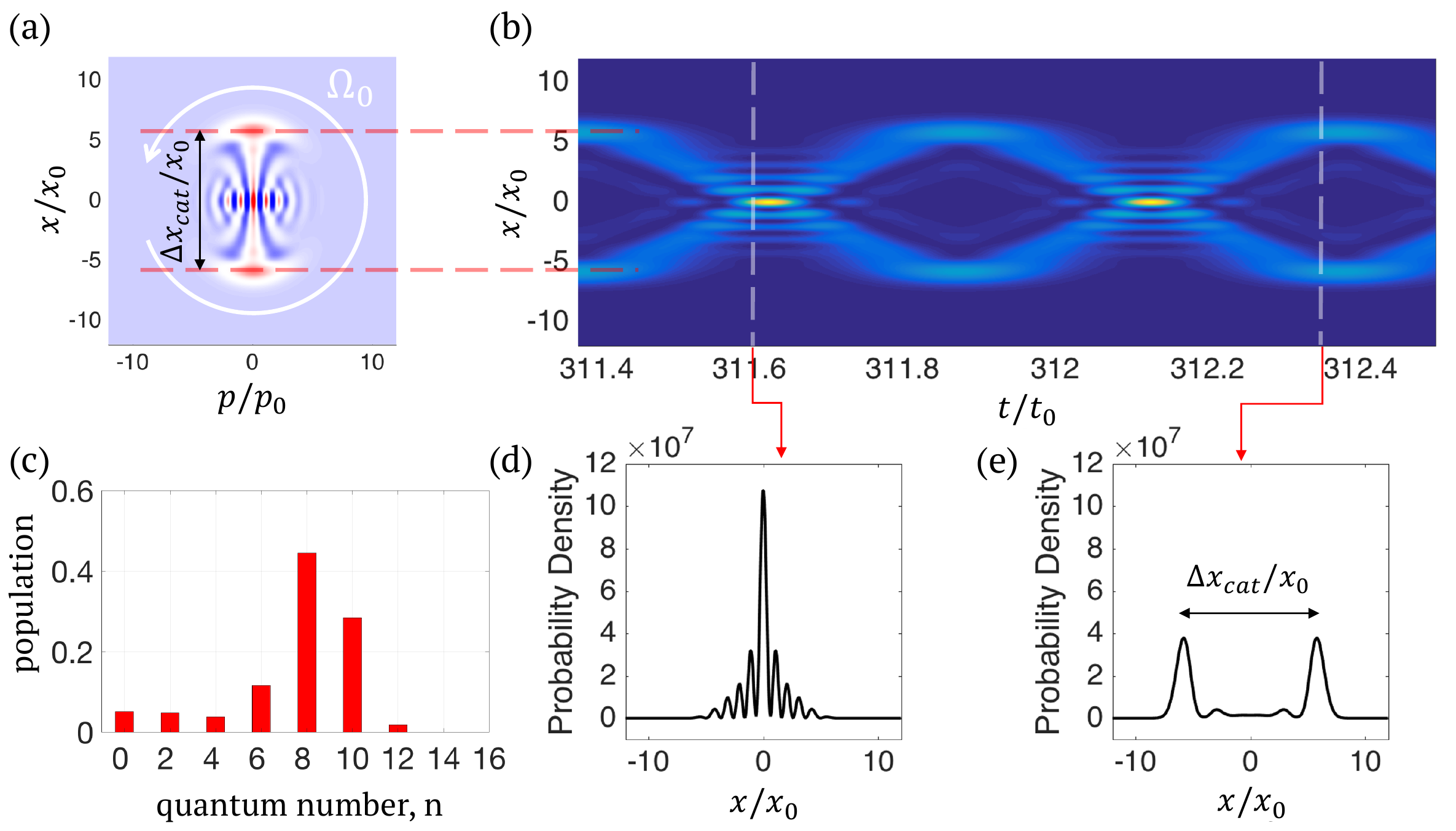}}
\caption{(color online) Free oscillation of an amplitude-squeezed Schr\"{o}dinger cat state in a harmonic potential. (a) The Wigner function $W(x,p,t)$ of an amplitude-squeezed Schr\"{o}dinger cat state rotates in the counter-clockwise direction as time evolves. Note that $p_{0} \equiv \hbar\ko$ and $t_{0} \equiv 2\pi/\wo$. (b) The probability distribution as a function of time is given by the projection of the Wigner function along the $x$-axis, $P(x,t)= \int_{-\infty}^{\infty} dp W(x,p,t)$. The two branches of the probability trace show that the two wavepackets of the cat state oscillate back and forth in the harmonic potential. (c) The population distribution of the amplitude-squeezed Schr\"{o}dinger cat state peaks at $\nmax = 8$. (d) The probability distribution shows interference fringes as the two wavepackets cross the potential minimum. (e) As the two wavepackets reach the turning points, the probability distribution shows two distinct peaks separated by $\Dx \approx 11\xo$.}
\label{fig:cat}
\end{figure}

While the above analysis assumes a cat state made of two coherent state wavepackets, the result can be used for other cat states as an approximation if the component wavepackets are Gaussian-like and the width of the population distribution is sub-Poissonian, i.e. amplitude squeezed. Assuming $\nmax \gg 1$, the energy expectation value of such cat states can be estimated as
\begin{equation}\label{energycat5}
	\avg{E}_{cat} \approx \nmax\hbar\wo,
\end{equation}
because the population distribution is narrowly centered around $\nmax$ with a sub-Poissonian width. Also, for wavepacket separation much larger than the squeezed widths $\Dx \gg \sigma_{x}$ and $\Dp \gg \sigma_{p}$, \Eq{energycat1} can be approximated as 
\begin{equation}\label{energycat6}
	\avg{E}_{1,2} \approx \frac{1}{2}M\wo^2\xc^2 = \frac{1}{2}M\wo^2\left( \frac{\Dx}{2}\right)^2
\end{equation}
at the turning points, and 
\begin{equation}\label{energycat7}
	\avg{E}_{1,2} \approx  \frac{\pc^2}{2M}  = \frac{1}{2M}\left( \frac{\Dp}{2}\right)^2
\end{equation}
at the potential minimum. Therefore, using \EqsThree{energycat5}{energycat6}{energycat7} we again obtain the relations between $\Dx$, $\Dp$, and $\nmax$, 
\begin{equation}\label{maxwidth}
	\frac{\Dx}{\xo} = \frac{\Dp}{\hbar\ko} \approx 4\sqrt{\nmax}.
\end{equation}
An example is given in \Fig{fig:cat}, which is an amplitude-squeezed Schr\"{o}dinger cat state prepared through our simulation of inelastic KD-effect. In this example, we see that $\nmax = 8$ and $\Dx/\xo = \Dp/\hbar\ko \approx 11$, which is consistent with \Fig{fig:cat}(a) and (e). In FIG. 6 of the main text, we show three very large cat states whose population distributions peak at $\nmax = 648$ (electron), 5348 (molecule) and 5368 (nanoparticle). The full width of the distributions are 6 (electron), 32 (molecule), and 22 (nanoparticle), which are well below the full Poissonian width $2\sigma_{_\textrm{P}} = 2\sqrt{\nmax}$. Therefore, the estimated maximum separations are $\Dx/\xo = \Dp/\hbar\ko \approx$ 102 (electron), 293 (molecule), 293 (nanoparticle), which are in good agreement with the probability distributions shown in FIG. 6 of the main text.

\subsection{IV. Determination of the highest eigenstate available for excitation}

\subsubsection{(a) Pondermotive trap for electrons}

We consider an electron in the ponderomotive trap described in \Eq{electrontrap1}, $U_{p}(x) = U_{0}\cos^2{(\ktrap x)}$. Expanding around the potential minimum $x_{m} = \Ltrap/4$ gives
\begin{equation}
	U_{p}(x) \approx U_{0}\ktrap^2\left[ x^2-\frac{1}{3}\ktrap^2 x^4 \right], 
\end{equation}
where $x$ is repurposed as the displacement from the minimum. The harmonic approximation, $U_{p}(x) \approx U_{0}\ktrap^2 x^2$, is valid when $x \ll \xmax$, where $\xmax = \sqrt{3}/\ktrap$. As the iterative excitation cannot continue effectively in the anharmonic regime, there is a limit to the maximum spatial separation of cat states, i.e. $\Dx \ll 2\xmax$. Using \Eq{maxwidth}, we identify the upper bound for $\nmax$ as,
\begin{equation}
	\nmax \ll \frac{3}{16\pi^2}\left( \frac{\Ltrap}{\xo} \right)^2.
\end{equation}

\subsubsection{(b) Dipole trap for polarizable particles}

We consider a polarizable neutral particle in the dipole trap described in \Eq{moleculetrap1},
\begin{equation}
	U_{d}(x) = - \frac{\alpha\IS}{2\epsilon_{0} c}\cos^2{(\ktrap x)} . 
\end{equation}
Close to the potential minimum $x_{m}=0$, the dipole potential can be approximated as a harmonic trap
\begin{equation}\label{dipoletrap}
	U_{d}(x) \approx \frac{\alpha\IS}{2\epsilon_{0} c}\ktrap^2\left[ x^2-\frac{1}{3}\ktrap^2 x^4 \right],
\end{equation}
where a constant in the expansion is dropped. The harmonic approximation is valid when $x \ll \xmax$, where $\xmax = \sqrt{3}/\ktrap$. As before, $\Dx \ll 2\xmax$. Using \Eq{maxwidth}, the upper bound for $\nmax$ is
\begin{equation}
	\nmax \ll \frac{3}{16\pi^2}\left( \frac{\Ltrap}{\xo} \right)^2.
\end{equation}